\documentclass[11pt]{article}
\usepackage{times}
\usepackage{epsfig}

\makeatletter

\DeclareSymbolFont{boldmath}{OML}{cmm}{b}{it}
\DeclareSymbolFontAlphabet{\mathb}{boldmath}
\DeclareMathAlphabet{\Bbb}{U}{msb}{m}{n}
\DeclareMathAlphabet{\euf}{U}{euf}{b}{n}

\DeclareMathSymbol{\balpha}{0}{boldmath}{"0B}
\DeclareMathSymbol{\bbeta}{0}{boldmath}{"0C}
\DeclareMathSymbol{\bgamma}{0}{boldmath}{"0D}
\DeclareMathSymbol{\bdelta}{0}{boldmath}{"0E}
\DeclareMathSymbol{\bepsilon}{0}{boldmath}{"0F}
\DeclareMathSymbol{\bzeta}{0}{boldmath}{"10}
\DeclareMathSymbol{\bfeta}{0}{boldmath}{"11}
\DeclareMathSymbol{\btheta}{0}{boldmath}{"12}
\DeclareMathSymbol{\biota}{0}{boldmath}{"13}
\DeclareMathSymbol{\bkappa}{0}{boldmath}{"14}
\DeclareMathSymbol{\blambda}{0}{boldmath}{"15}
\DeclareMathSymbol{\bmu}{0}{boldmath}{"16}
\DeclareMathSymbol{\bnu}{0}{boldmath}{"17}
\DeclareMathSymbol{\bxi}{0}{boldmath}{"18}
\DeclareMathSymbol{\bpi}{0}{boldmath}{"19}
\DeclareMathSymbol{\brho}{0}{boldmath}{"1A}
\DeclareMathSymbol{\bsigma}{0}{boldmath}{"1B}
\DeclareMathSymbol{\btau}{0}{boldmath}{"1C}
\DeclareMathSymbol{\bupsilon}{0}{boldmath}{"1D}
\DeclareMathSymbol{\bphi}{0}{boldmath}{"1E}
\DeclareMathSymbol{\bchi}{0}{boldmath}{"1F}
\DeclareMathSymbol{\bpsi}{0}{boldmath}{"20}
\DeclareMathSymbol{\bomega}{0}{boldmath}{"21}
\DeclareMathSymbol{\beps}{0}{boldmath}{"22}
\DeclareMathSymbol{\bthet}{0}{boldmath}{"23}
\DeclareMathSymbol{\bomeg}{0}{boldmath}{"24}
\DeclareMathSymbol{\bvphi}{0}{boldmath}{"27}

\@addtoreset{equation}{section}
\newcommand{\nwl}{\nonumber\\}
\newcommand{\sref}[1]{section~\ref{#1}}

\newcommand{\fref}[1]{figure~\ref{#1}}
\newcommand{\Fref}[1]{Figure~\ref{#1}}
\newcommand{\eref}[1]{(\ref{#1})}

\newcommand{\ft}[2]{{\textstyle{{#1}\over{#2}}}}
\newcommand{\dd}{\mathrm{d}} 
\newcommand{\ii}{\mathrm{i}}
\newcommand{\Tr}{\mathop{\mathrm{Tr}}}
\newcommand{\Trr}[1]{\Tr(#1)}

\newcommand{\grpSL}{\mathsf{SL}}
\newcommand{\algsl}{\euf{sl}}
\newcommand{\ZZ}{\Bbb{Z}}
\newcommand{\RR}{\Bbb{R}}
\newcommand{\ads}{\mathcal{S}}
\newcommand{\scri}{\mathcal{J}}

\newcommand{\one}{\mathbf{1}}
\newcommand{\ga}{\bgamma}
\newcommand{\om}{\bomega}

\newcommand{\diag}{\mathop{\mathrm{diag}}}
\newcommand{\cs}{\mathop{\mathrm{cs}}}
\newcommand{\sn}{\mathop{\mathrm{sn}}}
\newcommand{\arctanh}{\mathop{\mathrm{arctanh}}}

\newcommand{\eps}{\varepsilon}

\newcommand{\xx}{\mathb{x}}
\newcommand{\yy}{\mathb{y}}
\newcommand{\zz}{\mathb{z}}
\newcommand{\gh}{\mathb{g}}
\newcommand{\hh}{\mathb{h}}
\newcommand{\nn}{\mathb{n}}
\newcommand{\uu}{\mathb{u}}
\newcommand{\pp}{\mathb{p}}
\newcommand{\ww}{\mathb{w}}
\newcommand{\vv}{\mathb{v}}

\newcommand{\ch}{\chi}
\newcommand{\ph}{\varphi}
\newcommand{\cen}{\alpha}
\newcommand{\rad}{\beta}
\newcommand{\dir}{\theta}
\newcommand{\vh}{\xi}
\newcommand{\xh}{\zeta}
\newcommand{\eh}{\epsilon}

\newcommand{\mass}{m}
\newcommand{\imass}{\mu}

\newcommand{\wdg}{w}
\newcommand{\vdg}{v}

\makeatother

\begin{document}

\begin{flushright}
MZ-TH/98-55\\
gr-qc/9809087
\end{flushright}

\begin{center}
  \LARGE \textbf{Black hole creation in 2+1 dimensions}
\end{center}
 
\vspace*{8mm}

\begin{center}
  \textbf{Hans-J\"urgen Matschull}\\[2ex]
  Institut f\"ur Physik, Johannes Gutenberg-Universit\"at\\[1ex]
  Staudingerweg 7, 55099 Mainz, Germany\\[1ex]
  E-mail: matschul@thep.physik.uni.mainz.de
\end{center}

\vspace*{10mm}

\begin{abstract}
  When two point particles, coupled to three dimensional gravity with a
  negative cosmological constant, approach each other with a
  sufficiently large center of mass energy, then a BTZ black hole is
  created. An explicit solution to the Einstein equations is presented,
  describing the collapse of two massless particles into a non-rotating
  black hole. Some general arguments imply that massive particles can be
  used as well, and the creation of a rotating black hole is also
  possible.
\end{abstract}

\vspace*{10mm}

\section*{Outline}

The three dimensional black hole of Banados, Teitelboim and Zanelli
\cite{btz,bhtz} has turned out to be a useful toy model to study various
aspects of black hole quantum physics and thermodynamics. It is a
solution to the vacuum Einstein equations with a negative cosmological
constant. In its maximally extended version, its global structure is
very similar to the maximally extended Schwarzschild black hole, or
wormhole solution to Einstein gravity in four dimensions. Spacetime
splits into four regions, the interiors of a black hole and a white
hole, and two causally disconnected external regions. They are
asymptotically flat in the Schwarzschild case, whereas for the BTZ black
hole they are anti-de-Sitter spaces.

For the Schwarzschild black hole it is well know that it can be created
by, for example, a star collapse. What is necessary for such a collapse
is that a sufficiently large amount of matter is concentrated inside a
small region of space. If the black hole is created in this way, then
 only two of the four regions of spacetime exist: one exterior region,
which is asymptotically flat and contains the initial matter
configuration, and the interior of the black hole, which is separated
from the exterior by a future horizon. There is no white hole and no
second asymptotically flat region. In this sense, the star collapse is
more realistic than the wormhole solution, because it can evolve from a
singularity free initial condition.

Another way to create a Schwarzschild black hole is to start from a
collapsing spherically symmetric dust shell, which is somewhat easier to
deal with than a star, because there are no matter interactions other
than the gravitational ones. The analog solution to the three
dimensional Einstein equations, describing a circular dust shell
collapsing into a BTZ black hole, has in fact been found shortly after
the discovery of the BTZ black hole itself \cite{mannross}. Here, I
would like to present another way to create a BTZ black hole, starting
form a very different initial condition. Instead of a dust shell, which
can be considered as a special arrangement of infinitely many particles,
it is sufficient to consider just two particles, which approach each
other such that at some time they collide. The four dimensional analog 
of this process would be the collision of two stars, with sufficient 
masses and center of mass energy to create a black hole. 

The situation is however much simpler in three dimensions, because the
particles can be taken to be pointlike, and we can even choose them to
be massless, which simplifies the construction of an explicit solution
to the Einstein equations even further. This is because pointlike
particles in three dimensional gravity are very easy to deal with.
Unlike in higher dimensions, they do not themselves form black holes.
Their gravitational fields are conical singularities located on their
world lines. Outside the matter sources, spacetime is just flat,
respectively constantly curved for a non-vanishing cosmological constant
\cite{pp3d}. Spacetimes containing only such point particles as matter
sources can be constructed by cutting out special subsets, sometimes
called wedges, from Minkowski space, and then identifying the boundaries
of these subsets in a certain way \cite{thooft,matwel}.

After setting up the notation and summarizing some basic features of
anti-de-Sitter space in, I will give a brief description of this cutting
and gluing procedure and its generalization to anti-de-Sitter space. It
is then straightforward to consider a special process where two
particles collide and join into a single particle. It turns out that,
depending on the energy of the incoming particles, the joint object is
either a massive particle or a black hole. More precise, if the center
of mass energy of the incoming particles lies beyond a certain
threshold, then the object that is created after the collision is not a
massive particle moving on a timelike geodesic, but some other kind of
singular object, which is located on a spacelike geodesic.

A closer analysis of the causal structure of the resulting spacetime
shows that this object is the future singularity inside a black hole.
The black hole has all the typical features such as, for example, an
interior region which is causally disconnected from spatial infinity,
and there is also a horizon, whose size is a function of the amount of
matter that has fallen in. Finally, reconsidering the same process in a
different coordinate system will show that the black hole created by the
collapse is indeed the BTZ black hole.

\section{Anti-de-Sitter Space}
\label{s-ads}
Three dimensional anti-de-Sitter space $\ads$ can be covered by a
global, cylindrical coordinate system $(t,\ch,\ph)$, with a real time
coordinate $t$, a radial coordinate $\ch\ge0$, and an angular coordinate
$\ph$ with period $2\pi$, which is redundant at $\ch=0$. The metric is
\begin{equation}\label{ads-metric-ch}
 \dd s^2 = \dd\ch^2 + {\sinh^2}\ch \, \dd\ph^2  - {\cosh^2\ch} \, \dd t^2 . 
\end{equation}
It is useful to replace the radial coordinate $\ch$ by $r=\tanh(\ch/2)$,
which ranges from zero to one only. Anti-de-Sitter space is then
represented by an infinitely long cylinder of radius one in $\RR^3$.
Expressed in terms of the coordinates $(t,r,\ph)$, the metric becomes
\begin{equation}\label{ads-metric-rh}
 \dd s^2 =  \Big(\frac{2}{1-r^2}\Big)^2 \, 
    \big( \dd r^2 + r^2 \, \dd \ph^2 \big)
          - \Big(\frac{1+r^2}{1-r^2}\Big)^2 \, \dd t^2.
\end{equation}
The time $t$ will be considered as an ADM-like coordinate time,
providing a foliation of anti-de-Sitter space. The hyperbolic geometry
of a spatial surface of constant $t$ is that of the \emph{Poincar\'e
  disc}, which is conformally isometric to a disc of radius one in flat
$\RR^2$. Hence, anti-de-Sitter space can be considered as a Poincar\'e
disc evolving in time. The time evolution is however not homogeneous.
The lapse function, that is, the factor in front of the $\dd t$-term in
the metric, which relates the physical time to the coordinate time $t$,
depends on $r$. It diverges at the boundary of the disc, indicating that
the physical time is running infinitely fast there.

\subsection*{Geodesics}
\begin{figure}[t]
\begin{center}
\epsfbox{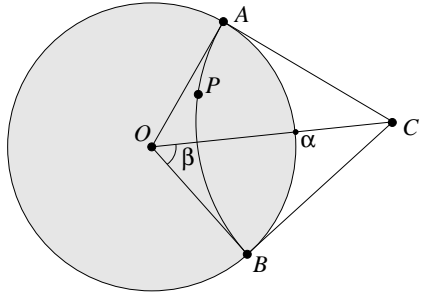}
\end{center}
\caption{Construction of a geodesic on the Poincar\'e disc}
\label{geo}
\end{figure}

The Poincar\'e disc has some nice properties, which allow a convenient
visualization of the constructions made in this article. The geodesics
on the disc are circle segments intersecting the boundary at $r=1$
perpendicularly. \Fref{geo} shows the construction of such a geodesic.
It is determined by two points $A$ and $B$ on the boundary. If the
angular coordinates of $A$ and $B$ are $\cen\pm\rad$, with $0<\rad<\pi$,
we call the circle segment $APB$ the geodesic \emph{centered} at $\cen$,
with \emph{radius} $\rad$. To derive an equation for the geodesic in
terms of the coordinates $r$ and $\ph$, consider the points in the
figure as complex numbers, such that $A=e^{\ii(\cen+\rad)}$ and
$B=e^{\ii(\cen-\rad)}$. It then follows that the center of the circle
segment $APB$ is at $C=e^{\ii\dir}/{\cos\rad}$, and that its radius is
$\tan\rad$. Using this, it is not difficult to show that, for a point
$P=r\,e^{\ii\ph}$ on the geodesic, we have
\begin{equation}\label{disc-geo}
   \frac{2\, r}{1+r^2} \, \cos ( \ph - \cen ) = \cos\rad.
\end{equation} 
This also applies for $\rad=\pi/2$, where the circle segment becomes a
diameter of the disc. In this case, the solution to \eref{disc-geo} is
$r=0$ or $\ph=\cen\pm\pi/2$, which is fulfilled on the diameter
orthogonal to the direction $\cen$.

Another special class of geodesics are spacetime geodesics passing
through the origin of the coordinate system at $r=0$ and $t=0$. They can
be parametrized by an angular direction $\dir$, and a velocity
$0\le\vh\le\infty$. The equation specifying such a geodesic in terms of
the cylindrical coordinates $(t,r,\ph)$ is
\begin{equation}\label{origin-geo}
   \frac{2\, r}{1+r^2}  = \vh \, \sin t, 
\end{equation}
and $\ph$ has to be equal to $\dir$. Let us consider this as a
definition of $r$ as a function of $t$, describing the motion of a test
particle in anti-de-Sitter space. To see that $\vh$ is its velocity when
it passes the origin, we have to differentiate \eref{origin-geo} at
$t=0$ and $r=0$. This gives $2\,\dd r=\vh\,\dd t$. As there is also a
relative factor of $2$ between $\dd r$ and $\dd t$ in the metric
\eref{ads-metric-rh}, the physical velocity is indeed equal to $\vh$.
For $0\le\vh<1$, the geodesic is timelike. The radial coordinate $r$ is
then oscillating with a period of $2\pi$ in $t$ between two extremals
$-\rho\le r\le\rho$, determined by
\begin{equation}\label{rh-max}
   \frac{2 \, \rho}{1 + \rho^2}  = \vh.
\end{equation}
The test particle starts off from the center of the disc at $t=0$,
moving into the direction $\ph=\dir$ with velocity $\vh$. At $t=\pi/2$,
it reaches the maximal distance, and it returns to the center at
$t=\pi$. Then it moves into the opposite direction, returns at $t=2\pi$,
and so on. To keep the equations describing this kind motion as simple
as possible, we have to allow negative values of $r$, with the obvious
identification of the point $(t,r,\ph)$ with $(t,-r,\ph\pm\pi)$.

For a lightlike geodesic with $\vh=1$, the relation \eref{origin-geo}
between $r$ and $t$ simplifies to $r = \tan(t/2)$, which holds for
$-\pi/2<t<\pi/2$. At the ends of this time interval, $r$ becomes equal
to one, which means that the geodesic reaches the boundary of the disc.
Unlike a timelike test particle, a light ray is not oscillating forth
and back. It travels once through the whole Poincar\'e disc, and the
amount of time that it takes to travel from one side to the other is
$\pi$. The existence of lightlike geodesics like this implies that the
discs of constant $t$ are not Cauchy surfaces of anti-de-Sitter space.
Causal curves enter at any moment of time from the boundary, and they
disappear there as well.

Being the origin and destination of light rays, the cylindrical boundary
of spacetime at $r=1$ is called $\scri$. To be precise, it is the
boundary of the conformal compactification of anti-de-Sitter space,
which is obtained by multiplying the metric with the conformal factor
$\ft14(1-r^2)^2$, which makes all coefficients of \eref{ads-metric-rh}
regular at $r=1$, and then including the boundary into the manifold. In
contrast to asymptotically flat spacetimes, we cannot separate the
origin of light rays $\scri_-$ from their destination $\scri_+$. There
are, for example, light rays emerging from $\scri$ \emph{after} other
light rays have arrived there.

Moreover, as a consequence of this unusual feature, $\scri$ itself has a
causal structure. The future light cone of a point on $\scri$ is the set
of all points where light rays emerging from that point arrive back on
$\scri$. To see how the light cones look like, it is sufficient to
consider the pullback of the conformal metric on the boundary at $r=1$,
which is given by $\dd\ph^2-\dd t^2$. The light rays on $\scri$ are thus
the left and right moving lines with $\dd t=\pm\dd\ph$. These are in fact
limits of light rays inside anti-de-Sitter space. In accordance with the
special light ray considered above, which passes through the origin of
anti-de-Sitter space, the amount of coordinate time that it takes for light
to travel from some point on $\scri$ to the opposite point is $\pi$. 

Finally, the geodesic equation \eref{origin-geo} also includes spacelike
geodesics. For $\vh>1$, it has a solution for $r$ only within an even
smaller time interval $-\tau<t<\tau$, where $\tau$ is determined by
$\sin\tau=1/\vh$. At the boundary of this interval, $r$ becomes equal to
one. In the limit $\vh\to\infty$, we also recover the diameters of the
Poincar\'e disc. Hence, a spacelike geodesics passes through the whole
disc even faster than a lightlike one, and it also starts and ends on
$\scri$. However, in contrast to the lightlike geodesic, which has a
vanishing physical length, its total length is infinite. In this sense,
the boundary $\scri$ can also be considered as spacelike infinity.

\subsection*{The group $\grpSL(2)$}
What we didn't show so far is that all these curves are actually
geodesics. To do this, it is convenient to use a different
representation of anti-de-Sitter space. It is isometric to the covering
of the group manifold $\grpSL(2)$, consisting of real
$2\times2$~matrices with unit determinant. As a basis of
$2\times2$~matrices, we introduce the unit matrix~$\one$, and the gamma
matrices
\begin{equation}\label{gamma}
 \ga_0 = \pmatrix{ 0 & 1 \cr -1 & 0 } , \qquad
 \ga_1 = \pmatrix{ 0 & 1 \cr 1 & 0} , \qquad
 \ga_2 = \pmatrix{ 1 & 0 \cr 0 & -1 }.
\end{equation}
They form an $\algsl(2)$ Clifford algebra,
\begin{equation}\label{gamma-alg}
  \ga_a \ga_b = \eta_{ab} \, \one - \eps_{ab}{}^c \, \ga_c, 
\end{equation}
where the indices $a,b,\dots$ run from $0$ to $2$,
$\eta_{ab}=\diag(-1,1,1)$ is the three dimensional Minkowski metric,
which is used to raise and lower indices, and $\eps^{abc}$ is the Levi
Civita symbol with $\eps^{012}=1$. Expanding a generic matrix $\xx$ in
this basis,
\begin{equation}\label{gamma-expand}
  \xx = x_3 \, \one + x^a \, \ga_a ,\qquad
   x_3 = \ft12\Trr{\xx} , \quad
   x^a = \ft12\Trr{\xx\,\ga^a},
\end{equation}
we obtain a scalar $x_3$ and a vector $x^a$. The condition for the
determinant to be one is
\begin{equation}\label{det-con}
  x_3^2 - x^a x_a = x_3^2 + x_0^2 - x_1^2 - x_2^2 = 1.
\end{equation}
This defines a unit hyperboloid in $\RR^{(2,2)}$. It is not simply
connected, because there is a non-contractible loop in the $(x_3,x_0)$
plane. To see that anti-de-Sitter space is the covering thereof, we
define a projection $\ads\to\grpSL(2)$. In terms of the coordinates
$(t,\ch,\ph)$ it is essentially the \emph{Euler angle} parametrization
of $\grpSL(2)$,
\begin{eqnarray}\label{ads-SL-ch}
  \xx &=& 
    e^{\frac12(t+\ph)\,\ga_0} \, e^{\ch\,\ga_1} 
     \, e^{\frac12(t-\ph)\,\ga_0} \nwl
      &=& \cosh\ch \, (\cos t \, \one   + \sin t \, \ga_0)
        + \sinh\ch \, (\cos\ph \, \ga_1 + \sin\ph \, \ga_2).
\end{eqnarray}
The projection is locally one-to-one, but not globally. The right hand
side of \eref{ads-SL-ch} is obviously periodic in $t$. The time
coordinate of anti-de-Sitter space is winded up on the group manifold,
with a period of $2\pi$. To check that the projection is an isometry,
one can show, by straightforward calculation, that the anti-de-Sitter
metric given above is equal to the pullback of the Cartan Killing metric
on $\grpSL(2)$, which is the same as the induced metric obtained by
embedding $\grpSL(2)$ into $\RR^{(2,2)}$,
\begin{equation}\label{SL-metric}
 \dd s^2 = \ft12 \Trr{ \xx^{-1} \dd \xx \, \xx^{-1} \dd \xx } 
         =  \dd x_1^2 + \dd x_2^2 - \dd x_0^2 - \dd x_3^2   .
\end{equation}
A compact way to write the projection in terms of the alternative
coordinates $(t,r,\ph)$ is
\begin{equation}\label{ads-SL-rh}
  \xx  = \frac{1+r^2}{1-r^2} \, \om(t)
       + \frac{2\, r}{1-r^2} \, \ga(\ph) ,
\end{equation} 
where 
\begin{equation}\label{ga-om}
  \ga(\alpha) = \cos\alpha \, \ga_1 + \sin\alpha \, \ga_2, \qquad
  \om(\alpha) = \cos\alpha \, \one   + \sin\alpha \, \ga_0.
\end{equation}
Together with their derivatives,
\begin{equation}\label{ga-om-der}
  \ga'(\alpha) = \cos\alpha \, \ga_2 - \sin\alpha \, \ga_1, \qquad
  \om'(\alpha) = \cos\alpha \, \ga_0 - \sin\alpha \, \one,
\end{equation}
they provide a rotated basis of $2\times2$ matrices. Some formulas for
products are
\begin{eqnarray}\label{ga-om-form}
       \ga(\alpha)  \, \ga(\beta) &=& \om(\alpha-\beta), \qquad
       \ga(\alpha)  \, \om(\beta)  = \ga(\alpha-\beta), \nwl
       \om(\alpha)  \, \om(\beta) &=& \om(\alpha+\beta), \qquad
       \om(\alpha)  \, \ga(\beta)  = \ga(\alpha+\beta).
\end{eqnarray}
It is now straightforward to show that the curves considered above are
geodesics of anti-de-Sitter space. What we have to show is that their
projections are geodesics on the group manifold. Let us first consider
the geodesics passing through the origin, that is, through the unit
element $\one\in\grpSL(2)$. On the group manifold, such geodesics are
one-dimensional subgroups, consisting of the elements
$\xx(s)=e^{s\,\nn}$, $s\in\RR$, where $\nn=n^a\,\ga_a$ is some vector in
the Lie algebra $\algsl(2)$. Let us take $\nn=\ga_0+\vh\,\ga(\dir)$, and
work out the exponential explicitly. This gives
\begin{equation}\label{origin-geo-par}
  \xx(s) = e^{s\,\nn} = 
     \cs\,s \, \one + \sn s \, ( \ga_0 + \vh  \, \ga(\dir) ),
\end{equation}
where the real and analytic functions $\sn$ and $\cs$ are defined such
that
\begin{eqnarray}
  \sn s     &=& \frac{\sinh(s\sqrt{\vh^2-1})}{s\sqrt{\vh^2-1}}
              = \frac{ \sin(s\sqrt{1-\vh^2})}{s\sqrt{1-\vh^2}}, \nwl
  \cs s     &=&       \cosh(s\sqrt{\vh^2-1}) 
              =        \cos(s\sqrt{1-\vh^2}).
\end{eqnarray}
Comparing this to \eref{ads-SL-rh}, we find the following relation
between the coordinates and the curve parameter $s$,
\begin{equation}
  \frac{1+r^2}{1-r^2} \, \cos t = \cs s, \qquad 
  \frac{1+r^2}{1-r^2} \, \sin t = \sn s, \qquad
  \frac{2\,r }{1-r^2}     = \vh \, \sn s, 
\end{equation}
and $\ph$ has to be equal to $\dir$. One of the three relations between
$r$ and $t$ turns out to be redundant, and eliminating the curve
parameter $s$, for example by taking the quotient of the last two
equations, we recover the relation \eref{origin-geo}.

\subsection*{Isometries}
To show that the circle segments \eref{disc-geo} are geodesics as well,
we can exploit the fact that, being a group manifold, anti-de-Sitter
space is maximally symmetric. Every geodesic can be obtained by acting
with an isometry on a geodesic passing through a special point. On a
simple group manifold, isometries are arbitrary combinations of left and
right multiplications with constants,
\begin{equation}\label{ads-iso}
  \xx \mapsto \gh^{-1} \xx \, \hh, \qquad
    \gh,\hh\in\grpSL(2).
\end{equation}
Except for time and space inversion, every isometry of $\grpSL(2)$ can
be written in this way, and the corresponding isometry of anti-de-Sitter
space is then determined up to a time shift $t\mapsto t + 2\pi z$,
$z\in\ZZ$. It follows that every geodesic on $\grpSL(2)$ can be written
as
\begin{equation}\label{gen-geo}
  \xx(s) = \gh^{-1} \, e^{s\,\nn} \, \hh, \qquad
      \gh,\hh\in\grpSL(2), \quad
      \nn \in \algsl(2).
\end{equation}
The parametrization is somewhat redundant, but for our purposes it is
sufficient. Choosing
\begin{equation}
   \gh = e^{-\frac12 \xh \, \ga(\cen)} \, e^{-\frac12 \tau \, \ga_0} , \quad
   \hh = e^{ \frac12 \xh \, \ga(\cen)} \, e^{ \frac12 \tau \, \ga_0} , \quad
   \nn = \ga'(\cen),
\end{equation}
a straightforward calculation gives
\begin{equation}
  \xx(s)     = \cosh\xh \, \cosh s \, \om(\tau) 
               + \sinh\xh \, \cosh s \,\ga(\cen)
               + \sinh s \, \ga'(\cen).
\end{equation}
If we compare this to \eref{ads-SL-rh}, we find that this is a curve
that entirely lies inside the Poincar\'e disc at $t=\tau$, with the
following relations between the coordinates $r$ and $\ph$ and the curve
parameter $s$,
\begin{eqnarray}
    \cosh\xh \, \cosh s  &=& \frac{1+r^2}{1-r^2} ,\qquad
                \sinh s   =  \frac{2\, r}{1-r^2} \, \sin(\ph-\cen), \nwl
    \sinh\xh \, \cosh s  &=& \frac{2\, r}{1-r^2} \, \cos(\ph-\cen).
\end{eqnarray}
Again, one of these equations is redundant, and after eliminating the
curve parameter $s$, we are left with a single relation between $r$ and
$\ph$,
\begin{equation}\label{plane-geo}
       \frac{2\, r}{1+r^2} \, \cos(\ph-\cen) = \tanh \xh,
\end{equation}
which is the same as \eref{disc-geo}, defining a geodesic centered at
$\cen$ with radius $\rad$, if we choose $\xh$ such that
$\cos\rad=\tanh\xh$.

\section{Point Particles}
\label{s-prt}
So far, we only considered empty anti-de-Sitter space, which is the
unique solution to the vacuum Einstein equations with a negative
cosmological constant, provided that the topology of spacetime is that
of $\RR^3$. Let us now include a matter source, in form of a pointlike
particle. The effect of such a particle is almost the same with or
without a cosmological constant. The most convenient way to construct a
spacetime containing a point particle is by \emph{cutting} and
\emph{gluing} \cite{thooft,matwel}.

\subsection*{A point particle in Minkowski space}
For a vanishing cosmological constant, the method works as follows. One
starts from flat Minkowski space, which is isomorphic to the Lie algebra
$\algsl(2)$, that is, the spinor representation of the three dimensional
Lorentz algebra. This is quite useful, because to adapt the procedure to
anti-de-Sitter space, we only need to replace the Lie algebra by the
group manifold. Let us introduce orthonormal coordinates $(t,x,y)$ on
Minkowski space, and write a general vector as
\begin{equation}
  \zz = t \, \ga_0 + x \, \ga_1 + y \, \ga_2.
\end{equation}
Now, consider a Lorentz transformation acting on $\zz$. It can be
written as the adjoint action of some group element $\uu$,
\begin{equation}\label{mnk-iso}
  \zz \mapsto \uu^{-1} \zz \, \uu ,\qquad
    \uu \in\grpSL(2).
\end{equation}
The fixed points of this map lie on the \emph{axis} of the Lorentz
transformation, consisting of all matrices $\zz$ that commute with
$\uu$. If we expand $\uu$ in terms of the gamma matrices, and define a
vector $\pp$ as
\begin{equation}
  \uu = u \, \one + p^a \, \ga_a , \qquad
  \pp = p^a \, \ga_a ,
\end{equation}
then the direction of $\pp$ specifies the axis. Let us assume that $\pp$
is timelike or lightlike. The axis can then be interpreted as the world
line of a particle, with momentum vector $\pp$. In three dimensional
Einstein gravity, the effect of such a particle as a matter source is
that it produces a conical singularity, whose \emph{holonomy} is $\uu$
\cite{matwel}. The resulting spacetime is everywhere flat, except on the
world line, and transporting a vector once around the world line results
in the Lorentz transformation \eref{mnk-iso}.

The complete spacetime can be constructed by cutting out a \emph{wedge}
from Minkowski space. The wedge is bounded by two half planes emerging
from the world line, such that one of them is mapped onto the other by
the given Lorentz transformation. Note that this requires the points on
the world line to be fixed. If we identify the two faces according to
the Lorentz transformation, we obtain a spacetime that is locally flat,
because the map that provides the identification is an isometry of
Minkowski space. There is however a curvature singularity on the world
line. By construction, it has the required property. Transporting a
vector around the particle results in a Lorentz transformation, which is
the same as the one that defines the identification.

\begin{figure}[t]
\begin{center}
\epsfbox{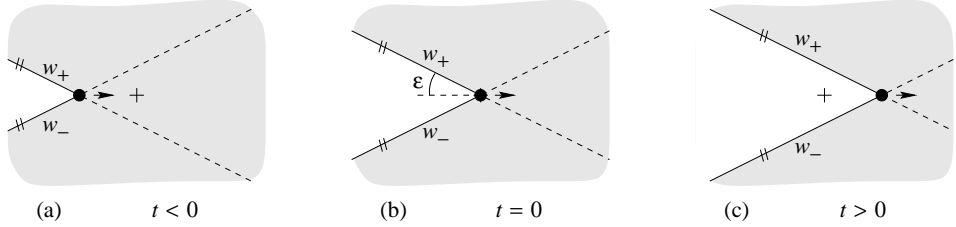}
\end{center}
\caption{A particle cutting out a wedge from Minkowski space.}
\label{fl1}
\end{figure}

A convenient way to visualize this construction is to use an ADM-like
foliation of spacetime. At a given moment of time $t$, the space
manifold is a plane with coordinates $x$ and $y$. From this plane, we
cut out a wedge, which is bounded by two half lines $\wdg_\pm$, such
that $\wdg_+$ is the Lorentz transformed image of $\wdg_-$. It is not
immediately clear that the wedge can be chosen like this. It is only
possible if the identification takes place within the planes of constant
$t$. It turns out that this can be achieved by choosing the wedge to lie
symmetrically in front of or behind the particle. For simplicity, let us
consider the following example. We choose a massless particle with a
lightlike momentum vector pointing into the $x$-direction. Its holonomy
is
\begin{equation}\label{mnk-hol}
  \uu = \one + \tan\eh \, (\ga_0 + \ga_1), \qquad
     0 < \eh < \pi/2.
\end{equation}
The corresponding isometry of Minkowski space is a lightlike, or
parabolic Lorentz transformation. The fixed points are at $x=t$ at
$y=0$. Hence, the particle is moving with the velocity of light from the
left to the right. To construct the wedge, we have to find a curve
$\wdg_-$ in the $t$-plane, such that its image $\wdg_+$ also lies in
this plane. Let us make the following symmetric ansatz. The world line
is invariant under vertical reflections, $y\mapsto-y$. So, we assume
that the wedge has this symmetry as well. A point $(t,x,y)\in\wdg_+$
then corresponds to the point $(t,x,-y)\in\wdg_-$. The matrix
representations of these points are
\begin{equation}
  \ww_\pm = t \, \ga_0 + x \, \ga_1 \pm y \, \ga_2,
\end{equation}
and for them to be mapped onto each other, we must have
$\uu\,\ww_+=\ww_-\,\uu$. Inserting the expressions for $\ww_\pm$ and
$\uu$, we find that this is fulfilled if and only if $y = (t-x) \,
\tan\eh$, which means that the faces $\wdg_\pm$ of the wedge are
determined by
\begin{equation}\label{wdg-mnk}
  \wdg_+ :\quad y =   ( t - x ) \, \tan\eh, \qquad
  \wdg_- :\quad y = - ( t - x ) \, \tan\eh.
\end{equation}
For a given value of $t$, these are two straight lines in the
$(x,y)$-plane, with angular directions $\pm\eh$. They intersect at the
fixed point, which is the position of the particle at time $t$.
\Fref{fl1} shows the lines $\wdg_+$ and $\wdg_-$ for three different
times $t$. The dot indicates the position of the particle, and the cross
represents that origin of the spatial plane at $x=0$ and $y=0$.

We can now cut out the wedge between the two lines, either in front of
or behind the particle. Both choices lead to the same spacetime
manifold, but covered with different coordinates. Let us choose the
wedge behind the particle. The space manifold is then the shaded region
shown in the figure, with the boundaries marked by the double strokes
identified. The opening angle $\eh$ of the wedge, which is half of the
\emph{deficit angle} of the conical space surrounding the particle, can
be considered as the total energy of the particle together with its
gravitational field, in units where Newton's constant is $G=1/4\pi$
\cite{matwel}. It is bounded from below by zero and from above by
$\pi/2$. Note that this energy is smaller than the energy of the
particle itself, that is, the zero component of its momentum vector,
which is $\tan\eh$.

The same construction can be made for massive particles. The holonomy
$\uu$ is then a timelike, or elliptic Lorentz transformation. It
represents a rotation of space around a timelike axis, which becomes the
world line of the particle. The wedge can be arranged in the same
symmetric way, and its opening angle is also equal to the total energy.
It is then bounded from below by the rest mass $\mass$, which is equal
to half of the angle of rotation. The angle of rotation coincides with
the deficit angle if the particle is at rest. The angle of rotation and
thus the rest mass of the particle can also be read off directly from
the holonomy $\uu$, using the \emph{mass shell} relation \cite{matwel},
\begin{equation}\label{mass-shell}
  \ft12\Trr{\uu} = \cos \mass, \qquad 
      0\le\mass\le\pi.
\end{equation}

\subsection*{A point particle in anti-de-Sitter space}
To construct a spacetime containing the same kind of point particle, but
with a negative cosmological constant, let us repeat the same procedure,
step by step, in anti-de-Sitter space. The first step was to specify the
world line as the set of fixed points of an isometry. We chose the
isometry to be a Lorentz transformation, which has the special property
that it leaves the origin of Minkowski space fixed. In principle, we
could weaken this restriction, but on the other hand we can always
choose coordinates such that the particle is passing through the origin.
It is therefore sufficient to consider those isometries of
anti-de-Sitter space which leave the origin fixed. We know already that
a general isometry of the group manifold $\grpSL(2)$ can be written as
\eref{ads-iso}. The unit element $\one$ is a fixed point if and only if
$\gh=\hh$. Hence, the isometry has to be of the form
\begin{equation}\label{prt-iso}
  \xx \mapsto \uu^{-1} \xx \, \uu, \qquad
  \uu \in \grpSL(2).
\end{equation}
So, the relevant isometry group is again $\grpSL(2)$. Indeed, $\uu$ can
be considered as the holonomy of the particle, in the same way as in
Minkowski space above. If we are in a neighbourhood of the origin, which
is small compared to the curvature radius of anti-de-Sitter space, we
can expand $\xx=\one+\zz+\dots$, with $\zz\in\algsl(2)$. On the
Minkowski vector $\zz$, the map \eref{prt-iso} acts exactly like the one
defined by \eref{mnk-iso} above. We can therefore expect that, in the
neighbourhood of the particle, spacetime will have the same conical
structure. Also in analogy with Minkowski space, the fixed points of the
given isometry are those elements of $\grpSL(2)$ that commute with
$\uu$. They can be found in the same way. We expand $\uu$ in terms of
the gamma matrices, and define a momentum vector $\pp$,
\begin{equation}\label{prt-mom}
  \uu = u \, \one + p^a \, \ga_a , \qquad
  \pp = p^a \, \ga_a.
\end{equation}
The only difference is that now the fixed points are not the vectors
proportional to $\pp$, but the elements of the one dimensional subgroup
generated by $\pp$, consisting of the matrices
\begin{equation}
  \xx(s) = e^{s \, \pp}, \qquad s \in \RR.
\end{equation}
This is a geodesic on the group manifold. Assuming that $\pp$ is
timelike or lightlike, it is the projection of a world line of a
massive, respectively massless particle in anti-de-Sitter space. As an
example, we consider the same massless particle once again, with
holonomy \eref{mnk-hol},
\begin{equation}\label{prt-hol}
  \uu = \one + \tan\eh \, ( \ga_0 + \ga_1 ).
\end{equation}
The fixed points lie on a lightlike world line, with $r = \tan(t/2)$ and
$\ph=0$. To construct the wedge that the particle cuts out from
anti-de-Sitter space, we proceed in the same way as before. First, we
switch to an ADM point of view, so that anti-de-Sitter space becomes a
space manifold, the Poincar\'e disc, evolving in time. Then, we look for
a pair of lines $\wdg_\pm$ on the disc of constant time $t$, which are
mapped onto each other by the given isometry. Finally, we cut out the
wedge between these lines, and identify the faces according to the
isometry.

There is however one crucial difference to the Minkowski space example
considered above. It only takes a finite amount of time for the particle
to travel through the whole disc. It enters at $t=-\pi/2$, and it leaves
again at $t=\pi/2$. Before and after that, there is no matter present,
and therefore spacetime is expected to be empty anti-de-Sitter space,
with no wedge or whatsoever cut out. Only for $-\pi/2<t<\pi/2$ the
particle is present, and we expect the space manifold to be a Poincar\'e
disc with a wedge cut out. For the shape of this wedge, we make the same
symmetric ansatz as in Minkowski space. The world line of the particle
is invariant under reflections of the vertical axis. In cylindrical
coordinates, this is the transformation $\ph\mapsto-\ph$. So, we assume
that a point $(t,r,-\ph)\in\wdg_-$ is mapped onto $(t,r,\ph)\in\wdg_+$.
The matrices representing these points on the group manifold are given
by \eref{ads-SL-rh},
\begin{equation}
  \ww_\pm = \frac{1+r^2}{1-r^2} \, \om(t) 
          + \frac{2\, r}{1-r^2} \, \ga(\pm\ph).
\end{equation}
Evaluating the equation $\uu\,\ww_+=\ww_-\uu$, we find that the faces
$\wdg_+$ and $\wdg_-$ are uniquely determined by the following
coordinate relations,
\begin{equation}\label{wedge-eq}
   \wdg_\pm:\quad \frac{2\, r}{1+r^2} \, \sin(\eh\pm\ph) 
          =  \sin t \, \sin \eh .
\end{equation}
If we define a parameter $\cen=\pi/2-\eh$, such that $\sin\eh=\cos\cen$,
and compare this to \eref{disc-geo}, we find that $\wdg_+$ is a geodesic
centered at $\cen$, whose radius $\rad$ is given by $\cos\rad = \sin t
\cos\cen$, and $\wdg_-$ is a geodesic with the same radius, centered at
$-\cen$. As a function of $t$, the radius decreases from $\rad =
\pi-\cen$ at $t =- \pi/2$ to $\rad = \cen$ at $t = \pi/2$.
\begin{figure}[t]
\begin{center}
\epsfbox{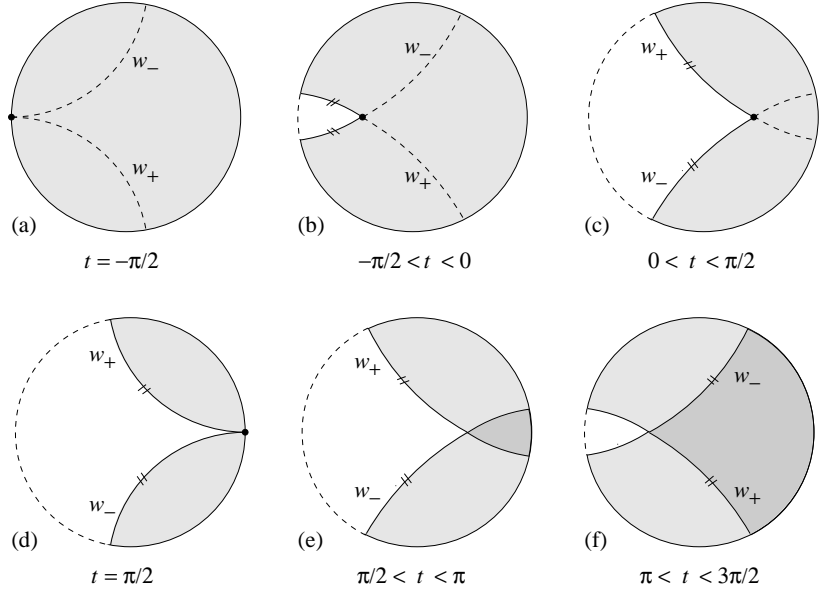}
\end{center}
\caption{A massless particle passing through anti-de-Sitter space} 
\label{pt1}
\end{figure}

In \fref{pt1} these curves are shown for different values of $t$. Let us
first consider the pictures (a--d), for the time between $t=-\pi/2$ and
$t=\pi/2$. What we see is that the lines $\wdg_+$ and $\wdg_-$ are
moving upwards, respectively downwards, such that their intersection,
the fixed point of the isometry at $r=\tan(t/2)$, moves from the left to
the right. The space manifold is obtained by cutting out the wedge
behind the particle and identifying the boundaries marked by the double
strokes. The resulting spacetime manifold has a constant curvature
everywhere, except on the world line. The reason is the same as before.
The map that provides the identification is an isometry of
anti-de-Sitter space, and therefore there is no extra curvature
introduced by gluing together the two faces of the wedge.

To see that the matter source is the same as before, it is, as already
mentioned, sufficient to consider a small neighbourhood of the world
line, where the curvature of anti-de-Sitter space can be neglected.
Indeed, if we enlarge the region around the center of the disc in
\fref{pt1}, it looks exactly like the one shown in \fref{fl1}. So, what
we have so far is a piece of spacetime between $t=-\pi/2$ and $t=\pi/2$.
The continuation has to be a solution to the vacuum Einstein equations,
because there is no matter present outside this time interval. This is
not in contradiction with causality, because the foliation of
anti-de-Sitter spacetime by discs of constant $t$ is not a foliation by
Cauchy surfaces. Matter consisting of massless particles can appear and
disappear at the boundary at any time.

For $t<-\pi/2$, it is more or less obvious how to continue. At
$t=-\pi/2$, the space manifold is a complete Poincar\'e disc, with no
matter inside. At this time the particle is still at $r=1$, which is
outside the open disc. Only a small time later the particle is actually
there. If we assume that for all earlier times the space manifold is a
complete disc as well, then we obtain a continuous solution to the
Einstein equations, which is matter free for all $t\le-\pi/2$. What is
not so obvious is what happens at $t=\pi/2$, after the particle has
left. The shaded region of \fref{pt1}(d) does not at all look like a
complete disc. Let us consider the further evolution of this space
manifold.

If we want to stick to $t$ as a global time coordinate, and the vacuum
Einstein equations to be fulfilled everywhere, then the time evolution
of the boundaries of the shaded regions is uniquely determined. This is
because the curves $\wdg_\pm$ defined by \eref{wedge-eq} are the only
curves inside the discs of constant $t$ which are mapped onto each other
by the given isometry. For times $t>\pi/2$, the curves start to move
backwards, as shown in \fref{pt1}(e--f), and then they oscillate between
two extremals. The shaded regions start to overlap, but this is just a
coordinate effect. One has to consider them as two separate charts
covering the space manifold. The only identification takes place along
the boundaries $\wdg_+$ and $\wdg_-$. As this is always defined by the
same isometry, we obtain a continuous solution to the Einstein equations
for all $t$, which is matter free outside the time interval
$-\pi/2<t<\pi/2$.

Now it seems that after the particle has left, spacetime looks very
different from what is was before, although we know that anti-de-Sitter
space is the only matter free solution to the Einstein equations on a
topologically trivial spacetime manifold. But this is also just a
coordinate effect. In fact, the foliation of spacetime by the space
manifolds shown in \fref{pt1}(e--f) is a somewhat skew foliation of
anti-de-Sitter space. To see this, let us take a three dimensional point
of view. The two shaded regions, evolving in time, then define two
subsets of anti-de-Sitter space, whose boundaries are the surfaces
$\wdg_\pm$. By definition, these two surfaces are mapped onto each other
by an isometry, which is continuous and one-to-one. Therefore, one of
the subsets is isometric to the complement of the other, and thus both
together form a complete anti-de-Sitter space.

So, finally we see that the whole situation is time symmetric. The
spacetime looks like empty anti-de-Sitter space before and after the
particle is there. The asymmetry in the pictures is only due to the fact
that it is not possible to cover the whole manifold symmetrically with a
single coordinate chart, which locally looks like the standard chart of
anti-de-Sitter space. We can reverse the picture if we cut out the wedge
in front of the particle instead of the one behind. We then obtain the
same spacetime, covered with different coordinates, providing the
standard foliation of anti-de-Sitter space after the particle has left,
but the skew one before it enters.

\section{Colliding Particles}
\label{s-col}
Let us now describe the process of two particles colliding, and thereby
joining and forming a single particle. The basic idea is as follows.
Consider two relativistic point particles in flat Minkowski space, with
no gravitational interaction. If they collide, that is, if their world
lines intersect at some point in spacetime, then we assume that from
that moment on they form a single particle, whose momentum vector is
given by the sum of the momenta of the incoming particles. This is
consistent with energy momentum conservation, and it is a deterministic
classical process, although not time-reversible. All properties of the
joint particle can be deduced from the incoming particles. In
particular, for a scalar particle the momentum vector is the only
relevant quantity.

When gravity is taken into account, the situation changes slightly. The
process is still deterministic, but it is not the sum of the momentum
vectors that is preserved. Instead, it is the total holonomy, which is
the product of the two holonomies of the incoming particles, and which
becomes the holonomy of the joint particle. This has some strange
consequences. For example, unlike the sum of two timelike or lightlike
vectors, the product of two timelike or lightlike holonomies is not
necessarily timelike. The joint particle can, for example, become a
tachyon. To understand this, it is again most convenient to study the
process in Minkowski space first, and then apply the same methods to
anti-de-Sitter space.

\subsection*{Joining particles in Minkowski space}
\begin{figure}[t]
\begin{center}
\epsfbox{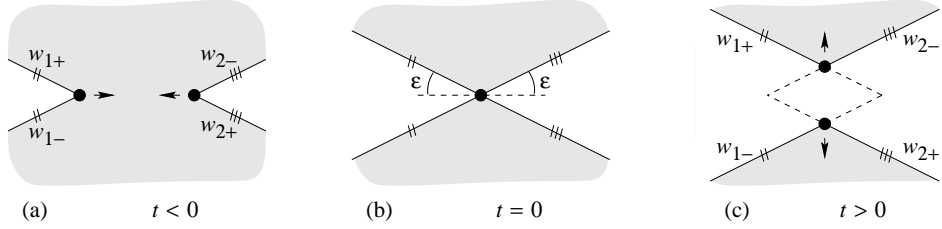}
\end{center}
\caption{Two particles colliding and joining.}
\label{fl2}
\end{figure}

Let us first consider the collision of two massless particles
graphically. We can always choose a coordinate system in Minkowski space
such that the collision takes place at the origin. The world lines of
both incoming particles, and also that of the outgoing particle, are
then passing through the origin, and we can apply the methods of the
previous section. Furthermore, we can choose a center of mass reference
frame, such that the particles come from opposite spatial directions and
have the same energy. Hence, without loss of generality, we can assume
that the holonomies of the incoming particles are given by
\begin{equation}\label{two-hol}
  \uu_1 = \one + \tan\eh \, ( \ga_0 + \ga_1 ) , \qquad
  \uu_2 = \one + \tan\eh \, ( \ga_0 - \ga_1 ).
\end{equation}
The first particle is then the same as the one considered in the
previous section. The wedge that it cuts out from Minkowski space is
bounded by the faces
\begin{equation}\label{wdg-mnk-1}
  \wdg_{1+} :\quad y =   ( t - x ) \, \tan\eh, \qquad
  \wdg_{1-} :\quad y = - ( t - x ) \, \tan\eh.
\end{equation}
The second particle has the same properties, except that it is moving into the opposite direction. The wedge is found by rotating that of the first particle by $180$ degrees,
\begin{equation}\label{wdg-mnk-2}
  \wdg_{2+} :\quad y = - ( t + x ) \, \tan\eh, \qquad
  \wdg_{2-} :\quad y =   ( t + x ) \, \tan\eh.
\end{equation}
For times $t<0$, before the collision, the space manifold is the shaded
region of \fref{fl2}(a). It is a plane with two wedges cut out behind
the two particles, both with opening angle $\eh$. The identification
along the boundaries, indicated by the double and triple strokes, is
again such that points with the same $x$-coordinate correspond to each
other. The resulting space manifold looks like a double cone, that is, a
cone with two tips, moving towards each other with the velocity of
light.

At $t=0$, when the particles collide, the space manifold becomes a
simple cone with a single tip. Assuming that this is also the case at
any later time, the further evolution of spacetime is uniquely
determined by the Einstein equations, without making any additional
assumptions about the joint particle itself. The argument is very
similar to the one used in the end of the previous section. It is
sufficient to know that spacetime is everywhere flat, except at one
point in space, which is the position of the joint particle. If we want
to stick to the foliation of spacetime by surfaces of constant $t$, then
there is only one way how the lines $\wdg_{1\pm}$ and $\wdg_{2\pm}$ can
evolve. They must be given by \eref{wedge-eq-1} and \eref{wedge-eq-2}
for all times, because these are the only lines inside the surface of
constant $t$, which are mapped onto each other by the given isometries.

In \fref{fl2}(c), we can see what they look like after the collision. If
we require that there is only one point in space where the curvature is
non-zero, then this can only be the point where $\wdg_{1+}$ and
$\wdg_{2-}$, respectively $\wdg_{2+}$ and $\wdg_{1-}$ intersect. Note
that due to the identification, these two points in the picture
represent the same physical point in space. The space manifold is then
covered by two charts, the two separate shaded regions, glued together
along their boundaries. The positions of the joint particle in the two
charts can be found as the intersections of the lines $\wdg_{1\pm}$ and
$\wdg_{2\pm}$. They lie on the $y$-axis, at $y={}\pm t \tan\eh$. Hence,
in the upper chart the particle is moving upwards with a velocity of
$\tan\eh$, and in the lower chart it is moving downwards with the same
velocity.

\subsection*{Energy momentum conservation}
As we should have expected, the velocity of the joint particle depends
on the energy of the incoming particles. What is however somewhat
peculiar is that, for sufficient high energies, the velocity becomes
bigger than one and thus the outgoing particle is moving faster than the
speed of light. In other words, two incoming massless particles with
sufficient high energy can form a tachyon. This is impossible for
relativistic point particles in Minkowski space, without gravitational
interaction. The sum of two lightlike momentum vectors is always a
timelike vector.

What makes the situation different if gravity is present, is, as already
mentioned, that it is not the sum of the momentum vectors of the
particles which is preserved, but the product of their holonomies. In
particular, the holonomy of the outgoing particle is given by the
product of the two holonomies of the incoming particles. To see this,
consider once again \fref{fl2}(c). If we transport a vector once around
the joint particle, then we have to pass once over the left wedge and
once over the right wedge. The result is that we have to act on the
vector first with the Lorentz transformation represented by $\uu_1$, and
then with the one represented by $\uu_2$. The holonomy of the joint
particle is therefore the product of two individual holonomies of the
incoming particles. We can multiply them in two different ways,
\begin{eqnarray}\label{join-hol}
  \uu_+     &=& \uu_2 \, \uu_1 = 
     (1-2\tan^2\eh)\, \one +
     2 \tan\eh \, (\ga_0 + \tan\eh \, \ga_2), \nwl
  \uu_-     &=& \uu_1 \, \uu_2 = 
     (1-2\tan^2\eh)\, \one +
     2 \tan\eh \, (\ga_0 - \tan\eh \, \ga_2).
\end{eqnarray}
There are two representations of the holonomy of the joint particle,
because spacetime is covered by two coordinate charts, the upper and
lower shaded region in the figure. Each expression represents the
holonomy in one of the charts. To find out which is which, consider the
momentum vectors
\begin{equation}
  \pp_+ = 2 \tan\eh \, (\ga_0 + \tan\eh \, \ga_2), \qquad
  \pp_- = 2 \tan\eh \, (\ga_0 - \tan\eh \, \ga_2).
\end{equation}
Obviously, $\pp_+$ describes a particle that is moving upwards with a
velocity of $\tan\eh$, and $\pp_-$ corresponds to a particle moving
downwards with the same velocity. This implies that the quantities with
a plus index are the ones corresponding to the upper chart. We can now
also see is that the joint momentum vector is timelike for
$0<\eh<\pi/4$. Only in this case, the resulting joint object can be
considered as a massive particle. Its mass is given by the formula
\eref{mass-shell}. Inserting either $\uu_+$ or $\uu_-$, we find that
\begin{equation}\label{join-mass}
  \sin(\mass/2) = \tan\eh.
\end{equation}
For small energies, this becomes $\mass\approx2\eh$, which is the naive
expression that applies when gravity is switched off. For higher
energies, it deviates from this flat space relation. At $\eh=\pi/4$, the
mass reaches its maximum $\mass=\pi$, and for $\eh>\pi/4$, the formal
solution for $\mass$ becomes imaginary, indicating once again that the
joint particle is a tachyon. What the existence of such a spacelike
conical singularity in spacetime means depends rather crucially on its
global structure. We shall therefore not go into more details regarding
the tachyon in Minkowski space, where it can be considered as a kind of
big bang or big crunch singularity \cite{review}. A more interesting
situation arises in anti-de-Sitter space, where the spacelike
singularity becomes the future singularity of a black hole.

\subsection*{Joining particles in anti-de-Sitter space}
\begin{figure}[t]
\begin{center}
  \epsfbox{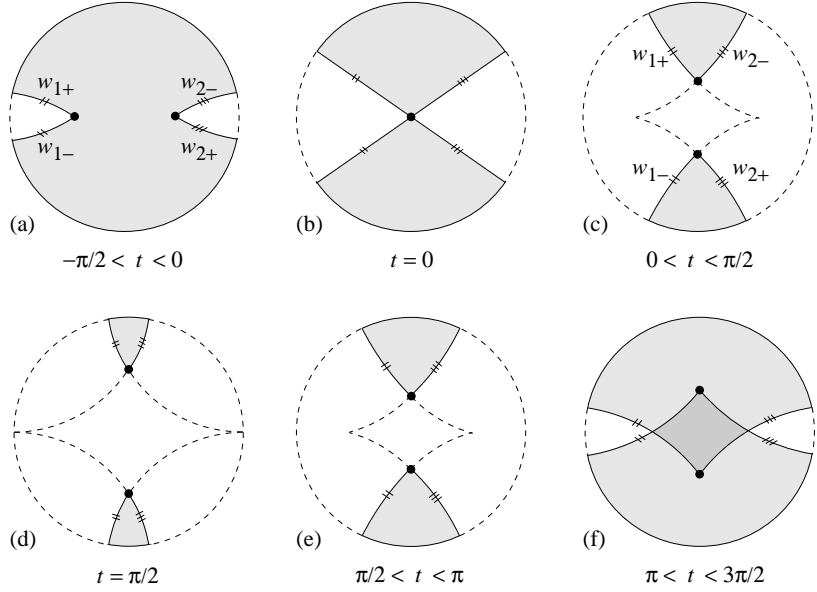}
\end{center}
\caption{Two massless particles joining and forming a massive particle.}
\label{pt2}
\end{figure}%
Before coming to this, let us first consider the colliding and joining
of two particles in anti-de-Sitter space, with the individual energies
being below the threshold, $0<\eh<\pi/4$. The construction is the same
as in Minkowski space, we only have to adapt the pictures and the
formulas. The holonomies of the incoming particles are given by the same
expressions \eref{two-hol}. Both world lines are lightlike geodesics
passing through the origin, entering from $\scri$ at $t=-\pi/2$. Before
that, the space manifold is a complete Poincar\'e disc. Afterwards, and
until the particles collide at $t=0$, it is a disc with two wedges cut
out, as shown in \fref{pt2}(a). The positions of the particles are on
the horizontal axis, at $r=\tan(t/2)$. The wedge of the first particle
is bounded by the curves \eref{wedge-eq},
\begin{equation}\label{wedge-eq-1}
   \wdg_{1\pm}:\quad \frac{2\, r}{1+r^2} \, \sin(\eh\pm\ph) 
             =  \sin t \, \sin \eh . 
\end{equation}
and that of the second particle is obtained by rotation,
$\ph\mapsto\ph+\pi$,
\begin{equation}\label{wedge-eq-2}
   \wdg_{2\pm}:\quad \frac{2\, r}{1+r^2} \, \sin(\eh\pm\ph) 
             =  - \sin t \, \sin \eh . 
\end{equation}
Locally, the process at $t=0$ is exactly the same as in Minkowski space.
If we enlarge a small neighbourhood of the origin of the disc at
$t\approx0$, everything looks the same as in \fref{fl2}. When the
particles collide in \fref{pt2}(b), the two conical singularities form a
single one. A short time later, the two shaded regions of \fref{pt2}(c)
separate from each other. Space is then covered by two coordinate
charts, and the positions of the joint particle in these charts can be
found as the intersection of the curves $\wdg_{1+}$ and $\wdg_{2-}$,
respectively $\wdg_{1-}$ and $\wdg_{2+}$. They lie on the vertical
diameter of the disc, with the radial coordinate given by
\begin{equation}\label{dot-ads}
  \frac{2\, r}{1+r^2} =   \sin t \, \tan\eh.
\end{equation}
This is a geodesic of the form \eref{origin-geo}. Whether it is
timelike, lightlike, or spacelike, and thus whether the joint particle
is slower or faster than the speed of light, depends on the size of
$\eh$. In fact, the relation between the velocity of the outgoing
particle and the energy of the incoming ones is the same as in flat
space. In accordance with the properties of the holonomy of the joint
particle \eref{join-hol}, it is timelike for $0<\eh<\pi/4$, lightlike
for $\eh=\pi/4$, and spacelike for $\pi/4<\eh<\pi/2$.

In case of a timelike particle, the further time evolution looks as
follows. The radial coordinate of the joint particle reaches a maximum
and the size of the shaded region has a minimum in \fref{pt2}(d), at
$t=\pi/2$. After that, the shaded regions of \fref{pt2}(e) are growing
again and moving towards each other. At $t=\pi$, the particle is back at
$r=0$, and thereafter the shaded regions start to overlap, as can be
seen in \fref{pt2}(f). This is however, once again, only a coordinate
effect. The space manifold is always covered by two separate charts,
with the only identifications taking place along the boundaries. The
maximal size of the space manifold is reached at $t=3\pi/2$, then it
starts to shrink again, and finally it oscillates between the two
extremals with a period of $2\pi$ in $t$. In contrast to a massless
particle, the massive particle does not leave the disc. This is because
only lightlike and spacelike geodesics can reach the boundary $\scri$,
but not timelike ones.

What is not so obvious is that the final state is indeed a spacetime
containing a single particle whose mass in given by \eref{join-mass}. In
the rest frame of such a particle, the complete spacetime would be an
anti-de-Sitter space, with the particle sitting in the center of the
Poincar\'e disc, from where a wedge with opening angle $\mass$ is cut
out. With a similar argument as in the end of \sref{s-prt}, one can show
that the foliation obtained here is again a somewhat skew foliation of
the standard spacetime where the particle is at rest. We are not going
to show this explicitly, because we are more interested in the case
where the energy of the infalling particles is beyond the threshold, and
the joint particle becomes a tachyon.

\section{Black Hole Creation}
\label{s-blh}
\begin{figure}[t]
\begin{center}
\epsfbox{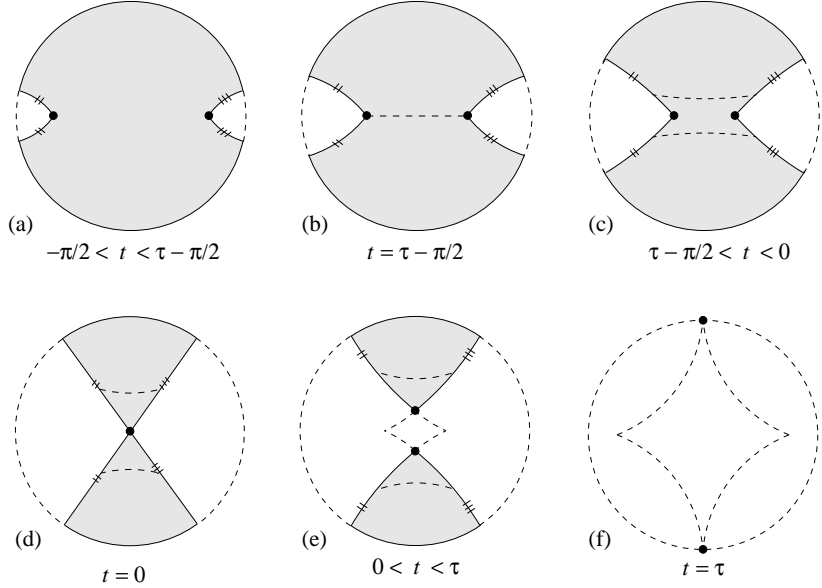}
\end{center}
\caption{The creation of a black hole}
\label{bh1}
\end{figure}

For $\pi/4<\eh<\pi/2$, the process of colliding and joining of the
particles, at least before and shortly after the collision, looks almost
the same as in \fref{pt2}. Only the parameter $\eh$, and thus the
opening angle of the two wedges, is somewhat bigger. The space manifolds
for three different times before the collision are shown in
\fref{bh1}(a--c), the collision takes place in \fref{bh1}(d), and
shortly after that the two shaded regions separate from each other. So
far, nothing new is happening. However, the dots in \fref{bh1}(e),
representing the joint object after the collision, are now moving
upwards, respectively downwards, on a \emph{spacelike} geodesic, which
is defined by
\begin{equation}\label{dot-bh}
  \frac{2\, r}{1+r^2} =   \sin t \, \tan\eh .
\end{equation}
As we already know from the general consideration of spacelike
geodesics, this equation has a solution only within a finite time
interval. The geodesic reaches the boundary of the Poincar\'e disc at
some time $t=\tau$, which is in this case given by
\begin{equation}\label{bh-tmax}
  \sin\tau = \cot\eh, \qquad
  0 < \tau < \pi/2.
\end{equation}
So far, this is not a problem. The massless particle in \fref{pt1} also
reaches the boundary of the disc and disappears at some time, but
nevertheless it was possible to continue spacetime afterwards. But now
the situation is different. If we look at \fref{bh1}(f), we see that,
together with the joint particle, the whole space manifold has
disappeared at $t=\tau$. There is nothing to be continued after that
moment. It is not immediately clear whether this is again just a
coordinate effect or something more relevant, but let us for the moment
assume that it is indeed not possible to continue. Then we have to
conclude that the spacelike world line of the joint object is a
\emph{future singularity}, that is, a line on which time ends.

\subsection*{The horizon}
If there is such a singularity in spacetime, then it is, in addition to
$\scri$, another possible destination of causal curves. We can then ask
the question whether there is a region of spacetime, from where every
causal curve ends on the singularity. Or, in other words, a region which
is causally disconnected from $\scri$, in the sense that no information
can be passed from there to spatial infinity. The boundary of such a
region is usually called a horizon, and what is beyond the horizon is
the interior of a black hole.

Before showing that there is such a horizon, let us first consider
$\scri$ itself. In \fref{bh1}, it is that part of the boundary of the
disc, which is also a boundary of the shaded region. For $t\le-\pi/2$,
this is the full boundary of the Poincar\'e disc. For $-\pi/2<t<\tau$,
it consists of two parts, one between the curves $\wdg_{1+}$ and
$\wdg_{2-}$, and one between $\wdg_{2+}$ and $\wdg_{1-}$. Because the
end points of the two parts are identified, $\scri$ has always the shape
of a closed circle. However, the circumference of this circle shrinks,
and it goes to zero at $t=\tau$. Let us call the point which is
represented by the two dots in \fref{bh1}(f) the \emph{last point} on
$\scri$. Note again that the identification is such that both dots
represent the same physical point on $\scri$.

It is not difficult to show that every point on $\scri$ is causally
connected to the last point. This is because the end points of the
curves $\wdg_{1\pm}$ and $\wdg_{2\pm}$ on $\scri$ are moving slower than
the speed of light. The amount of time they need to traverse a quarter
of the circumference of the disc is $\tau+\pi/2$, whereas a light ray
only needs $\pi/2$ to travel over the same distance. From this we
conclude that the horizon is the backward light cone of the last point
on $\scri$. To construct this light cone, we make use of some general
properties of light cones emerging from $\scri$. First of all, such a
light cone is a geodesic surface. If we intersect it with another
geodesic surface, say, a disc of constant $t$, then the intersection is
a geodesic. Moreover, we know that light travels on $\scri$ with a
velocity of $\dd\ph/\dd t=1$.

Using all this, it follows that the backward light cone of the upper dot
in \fref{bh1}(f) is, at some time $t$ in the past, a geodesic centered
at $\pi/2$, with radius $\tau-t$. The backward light cone emerging from
the lower dot is centered at $-\pi/2$, and has the same radius. These
two curves are shown as the dashed lines inside the shaded regions of
\fref{bh1}. Due to the identifications along the boundaries of the
wedges, they form a closed circle. From the area that is enclosed by
this circle, no signal can be sent along a causal curve to $\scri$, even
if we make use of the identifications and jump over the wedges. On the
other hand, from every point outside the circle, we can find a causal
curve either to the upper or to the lower part of $\scri$. This is also
possible for every point in space before $t=\tau-\pi/2$, which is the
moment when the two dashed lines meet in the middle of the disc.

All together we get the following scenario. At $t=-\pi/2$, the particles
enter from opposite sides, approaching each other with the speed of
light. When they are close enough, at $t=\tau-\pi/2$, the horizon
emerges, and at the same time the particles fall behind it. The actual
collision of the particles, and the creation of the future singularity,
takes place behind the horizon at $t=0$. This is very similar to the
creation of a Schwarzschild black hole in a star collapse. When the star
becomes small enough, the horizon emerges from some kind of cusp, and
the singularity is created behind the horizon. The differences are
mainly due to the different symmetries of the process. For example, the
horizon does not emerge from a point in the center of the star, but from
the geodesic that connects the two particle at the moment when they fall
behind it.

Another and more crucial difference seems to be that time also ends at
$t=\tau$ in the exterior region, which is not the case for a
Schwarzschild black hole. However, it is only the \emph{coordinate} time
$t$ that ends at this point. The physical lifetime of an observer
staying outside the horizon can be infinite. This is because the lapse
factor in front of the $\dd t$-term in the metric \eref{ads-metric-rh}
diverges at the boundary of the disc. Consider an observer moving on a
timelike curve which stays outside the horizon. There is no such
geodesic, which is another difference to the Schwarzschild black hole,
but there are timelike curves approaching the last point, or any other
point on $\scri$. In general, such a curve has an infinite physical
length, which means that the outside observer can live infinitely long.
This is not the case for a curve ending on the future singularity. The
world line of an observer falling into the black hole always ends after
a finite physical time.

\subsection*{The size of the black hole}
All these features justify the notion of a black hole for the object
generated by the two particles. We can then ask the question how big
this black hole is. As we are in two space dimensions, the size is
specified by the circumference of the horizon. It is constant in time,
because there is no more matter falling into the black hole after it has
been created and the two particles have fallen in. We shall see this
more explicitly in the next section. At $t=\tau-\pi/2$, the horizon
length is twice the physical distance between the two particles. To
calculate this explicitly, let us introduce a new parameter $\imass>0$,
which is essentially the analytic continuation of the rest mass $\mass$
of the massive particle, which was given by \eref{join-mass} for
$\eh<\pi/4$. For $\eh>\pi/4$, we define $\imass$ such that
\begin{equation}\label{join-btz}
  \cosh(\imass/2) = \tan\eh.
\end{equation}
Now, let $\rho$ be the radial coordinate of the particles at the moment
when the horizon is created,
\begin{equation}
 \rho = \tan(\tau/2-\pi/4).
\end{equation}
The time $\tau$ was given by \eref{bh-tmax}. Using this and some
trigonometric identities, we obtain
\begin{equation}
 \rho = \tanh(\imass/4).
\end{equation}
To find the physical distance between the particles, we have to
integrate the $\dd r$-component of the metric,
\begin{equation}
  \int_{-\rho}^{\rho} \frac{2}{1-r^2} \, \dd r  
      = 4 \, \arctanh \, \rho = \imass.
\end{equation}
It follows that the length of the horizon is $2 \imass$. Assuming that
there is some kind of no hair theorem, this should be the only parameter
describing a non-rotating black hole. However, we already considered
another quantity to specify this particular solution to the Einstein
equations, namely the total holonomy \eref{join-hol}. So, there should
be a relation between the holonomy and the length of the horizon. From
\eref{join-btz}, we infer that
\begin{equation}
  \cosh \imass = {\tan^2}\eh - 1.
\end{equation}
This is, up to a sign, the contribution proportional to the unit matrix
of the holonomy $\uu_+$ or $\uu_-$ in \eref{join-hol}. They both lie in
the same conjugacy class of $\grpSL(2)$, which is specified by
\begin{equation}\label{btz-shell}
  \ft12\Trr{\uu} = - \cosh\imass. 
\end{equation}
Hence, the size of the black hole can be read off directly from its
holonomy. The remaining information contained in the holonomy $\uu$ is
not a property of the black hole itself. It describes its motion
relative to the reference frame. We should also note that the relation
\eref{btz-shell} between the holonomy and the size of the black hole is
essentially the analytic continuation of the mass shell relation
\eref{mass-shell}. Geometrically, the mass was defined to be half of the
angle of rotation of the Lorentz transformation obtained by transporting
a vector around the particle. Here, the Lorentz transformation
represented by the holonomy $\uu$ is not a rotation, but a boost, and
$\imass$ is half of the hyperbolic angle that parametrizes this boost.
Using this analogy, we shall call $\imass$ the \emph{mass} of the black
hole, which is equal to half of the horizon length.

\subsection*{Extremal black holes}
We have now studied the collision of two particles with energies below
and above the threshold of $\eh=\pi/4$, but we didn't consider the case
$\eh=\pi/4$ so far. What happens in \fref{bh1} in this limit? The joint
object is no longer moving on a spacelike, but on a lightlike geodesic.
The time when it reaches the boundary of the disc becomes $\tau=\pi/2$.
As a consequence, the horizon emerges at $t=0$, at the moment when the
particles collide. As both the horizon and the joint particle move with
the speed of light, there is actually no interior region. This is in
accordance with \eref{join-btz}, which says that for $\eh=\pi/4$ we have
$\imass=0$, and therefore the size of the black hole becomes zero. So,
what we get in the lightlike limit is a kind of \emph{extremal} black
hole, whose inside region consists of only the singularity, located on a
lightlike geodesic.

But what is the difference between this and a massless particle, which
is also moving on a lightlike geodesic? Both objects have a lightlike
holonomy, representing a parabolic Lorentz transformation, but
nevertheless there is a crucial difference. For the lightlike particle
considered in \sref{s-prt}, the moment when it reaches $\scri$ is not
the end of time. Here it is, because even in the lightlike limit, there
is no space manifold left at $t=\tau$. To understand the difference, let
us also consider the lightlike case as a limit of the timelike one,
which is shown in \fref{pt2}. The reason why this could be extended for
all times in the future was that, when the radial distance of the joint
particle from the origin reached its maximum at $t=\pi/2$, there was
still some space left which could evolve further.

This is no longer the case when the joint particle becomes lightlike,
because then it reaches the boundary at $t=\pi/2$. The lightlike limit
of \fref{pt2} is thus the same as the lightlike limit of \fref{bh1}.
But we can also see is that, when considered as a limit of the timelike
situation, the extremal black hole is not the limit where $\mass\to0$.
Instead, it is the limit $\mass\to\pi$. This is the upper bound for the
rest mass of a particle in three dimensional Einstein gravity, 
because the maximal deficit angle of a conical singularity is $2\pi$. A
peculiar feature of this upper bound is that the world line of a
particle with $\mass=\pi$ is lightlike, but the structure of the conical
singularity on the world line is very different from the one generated
by a particle with $\mass=0$.

These \emph{exotic} lightlike particles also exist in flat spacetime,
with vanishing cosmological constant, and they also have some peculiar
features, which are very similar to extremal black holes
\cite{matwel,review}. The difference between massless and exotic
lightlike particles can also be seen at the level of the holonomies.
There are two different classes of group elements $\uu\in\grpSL(2)$
representing parabolic Lorentz transformations, namely those with
$\ft12\Trr{\uu} = 1$, and those with $\ft12\Trr{\uu} = -1$. The former
correspond to massless particle, and the latter are the extremal black
holes.

\subsection*{Massive particles}
As a possible generalization, let us ask the question whether it is
really necessary to start with massless particles, or can the black hole
also be created by two colliding massive particles? The answer is that
it can, but then the global structure of spacetime changes, at least if
we do not allow any other interaction between the particles than the
gravitational one. If we want the same black hole to be created from two
infalling massive particles, then the only constraint is that the
product of their holonomies is the same, or at least that it lies in the
same conjugacy class of $\grpSL(2)$. This can be achieved for any given
pair of masses. The structure of spacetime after the collision is then
exactly the same as the one shown in \fref{bh1}(d--f). What is also
similar is that the horizon is created and the particles fall behind it
at some time $t=\tau-\pi/2$, where $\tau$ is the time coordinate of the
last point on $\scri$.

However, the situation looks very different before that. Unlike the
massless particles, the massive ones cannot enter from $\scri$. They
must have been there for all times in the past. At $t=-\pi/2$, the
timelike geodesics on which the particles are moving reach their maximal
distance from the center. If we follow their world lines back into the
past, then they meet again in the center at $t=-\pi$. So, something
special is happening there as well. In fact, the whole situation is
symmetric with respect to time reversion $t\mapsto-\pi/2-t$. Using the
same arguments as above for the collision of the particles, we have to
conclude that the only reasonable way to continue further to the past is
to assume that the particles emerge from the decay of some other object.
This joint object must be a past singularity, that is, a spacelike curve
on which time begins and causal curves cannot be extended to the past.
It reaches $\scri$ at the time $t=-\pi-\tau$, and there is no extension
of spacetime before that.

A foliation of this spacetime looks like the one shown in \fref{bh1},
but starting not from a full Poincar\'e disc for $t\le-\pi/2$, but from
the singular situation (f) at $t=-\pi-\tau$, then going backwards to (a)
at $t=-\pi/2$, and ending up again with the last picture (f) at
$t=\tau$. There is then also a \emph{first point} in $\scri$, and the
light cone emerging from this point is the past horizon of a \emph{white
  hole}. An outside observer in this spacetime sees two particles coming
out of a white hole, reaching a maximal distance, approaching each other
again, and finally falling into a black hole. This was the actual reason
for considering the collapse of two massless particles first. With that
kind of matter it is possible to give an exact solution to the Einstein
equations, describing the creation of a black hole, without the need of
any other exotic structure like the white hole.

We thereby exploited the global causal structure of anti-de-Sitter
space, and in particular that of the boundary $\scri$, which is very
different from that of asymptotically flat spacetimes. An alternative
way to avoid the white hole, but nevertheless create a black hole using
massive particles, might be to allow another kind of interaction, which
prevents the particles from approaching each other in the past. It is
not immediately clear whether this can be achieved, in particular
because, for a consistent treatment, one has to describe the second
force as a field which also interacts with gravity, and this makes it
much harder to find an exact solution. The only way to avoid the white
hole without introducing other forces is to create the massive particle
themselves, or at least one of them, out of two massless particles, by
the process described in \sref{s-prt}.

\subsection*{Rotating black holes}
Finally, let us very briefly consider another possible generalization.
What happens if the particles do not collide, but pass each other at,
say, a small distance? An intuitive conclusion would be that they create
a rotating black hole. But is there any singularity then? If there is no
collision, then there is also no need for the particles to join and to
form a single object. They are just scattered, move on along their
lightlike geodesics, and disappear to $\scri$ at some time in the
future, such that in the end, spacetime looks like anti-de-Sitter space
again. If we look at this process in Minkowski space, then we find
indeed that, if the particles do not hit each other, they just move on
along their world lines extending to infinity.

This is not immediately obvious from \fref{fl2}, because the two wedges
shown there will in any case overlap from some time on, and then we have
to switch to a different coordinate system. But if we give up the
condition that spacetime should be foliated by planes of constant $t$,
it is easy to see that, given any two non-intersecting lightlike world
lines in three dimensional Minkowski space, one can always choose the
wedges such that they do not overlap. If there is no joint object any
more, one can also ask the question whether there is still anything
special about the threshold $\eh=\pi/4$, which was previously the
minimal energy of the incoming particles to create a spacelike
singularity. In the case of colliding particles, it makes a crucial
difference whether the energy is below or above the threshold, but it
seems that, if the particles are just scattered, this does not play a
particular role.

But this is not true. There is a crucial difference between the two
situations, even if no joint object is created. The difference is that,
if the energy lies above the threshold, then there are closed timelike
curves in spacetime. These curves are always present, however large the
distance between the particles is, and they are not located in the
neighbourhood of the particles, but in a region of spacetime which
extends, from outside some area around the particles at the moment of
closest approach, to spatial infinity. For a vanishing cosmological
constant, this spacetime is known as the Gott universe
\cite{gott,hgott}. There is no singularity like the one for the
non-rotating black hole, but instead there is a region in spacetime
where closed timelike curves exist.

Such a region of spacetime behaves very much like a future or past
singularity. Causal curves can end, respectively start there, because
they can wind up in an endless loop. Indeed, the analogy between closed
timelike curves and future or past singularities is very close. Remember
that above we assumed that it is not possible to continue the black hole
spacetime beyond the future singularity. But this is only true if we do
not allow closed timelike curves. If we give up this restriction, then
there is a possible extension. It is very similar to the Misner
universe, which can be obtained by cutting out a piece from two
dimensional Minkowski space, and identifying the boundaries according to
a hyperbolic Lorentz transformation.

Turning this argument around, we should consider a region of spacetime
with closed timelike curves as a future, respectively past singularity
as well. Doing so, we find the following scenario of two massless
particles, passing each other with sufficiently high energy. They enter
the spacetime from $\scri$ at some time, and before that we have an
empty anti-de-Sitter space. The particles are then approaching each
other, and some time before the moment of closest approach, closed
timelike curves arise. They appear at spatial infinity first, and then
the boundary of the region where they exist, now interpreted as a
singularity in spacetime, approaches the particles. But it never hits
them, because there are no closed causal curves intersecting the world
lines. There remains a region of space around and between the particles
where no closed timelike curves ever occur.

After the moment of closest approach, the particles just move on, the
boundary of the region with closed timelike curves moves back to spatial
infinity, and the same happens to the particles, which disappear to
$\scri$ at some time in the future. After that, spacetime looks again
like empty anti-de-Sitter space. We can roughly divide it into three
parts, the time before the singularity appears, while it is there, and
after it has gone. The boundary $\scri$ of this spacetime splits into
two disconnected parts, before and after the singularity. Note that,
because $\scri$ is two dimensional, the existence of already a single
closed causal curve, interpreted as a singularity, implies that $\scri$
splits into two parts, because the curve cuts it into two pieces. The
singularity on $\scri$ the same as a \emph{last point}, respectively a
\emph{first point} on $\scri$, depending on which part of $\scri$ we are
looking at.

The future and past light cones of this singularity define two horizons,
which separate the three parts of spacetime from each other. The
backward light cone is the future horizon of a black hole sitting in the
lower region of spacetime, the forward light cone is the past horizon of
a white hole in the upper region, and the region between the two
horizons is a wormhole that connects the two exterior regions, which are
otherwise completely disconnected. This is the typical structure of a
timelike wormhole, except that usually the process is periodic in time.
But this can again be achieved by using massive particle instead of
massless ones. There will then be a whole series of wormholes connecting
a sequence of exterior universes, each with a lifetime of $\pi$ in the
time coordinate $t$.

All together, it seems that a Gott universe in anti-de-Sitter space is
almost the same as a rotating BTZ black hole, and studying a point
particle collapse might clarify some still open questions regarding the
rotating black hole in three dimensions \cite{adsbh,rotbh}. The
arguments given here were based on the assumption that the results from
Minkowski space can be straightforwardly generalized to anti-de-Sitter
space. In particular, we had to assume that the region of spacetime
containing closed timelike curves has the same structure. Actually, this
is only justified in a neighbourhood of the particles, and if they pass
each other at a small distance, as compared to the curvature radius of
anti-de-Sitter space. So, a more detailed analysis is necessary to
answer the question whether the relation is indeed that close.

\section{The BTZ Black Hole}
\label{s-btz}
The purpose of this last section is to show that the black hole created
after the collision of the particles is indeed the same as the BTZ black
hole. To do this, we shall consider the same process in a different
coordinate system. So far, we used a center of mass frame, in which the
sum of the spatial momenta of the two infalling particles was zero. Now,
we shall switch to a kind of rest frame of the black hole. This is not
the same, again because it is not the sum of the momenta but the product
of the holonomies that determines the motion of the black hole relative
to the reference frame.

\subsection*{The maximally extended black hole}
\begin{figure}[t]
\begin{center}
  \epsfbox{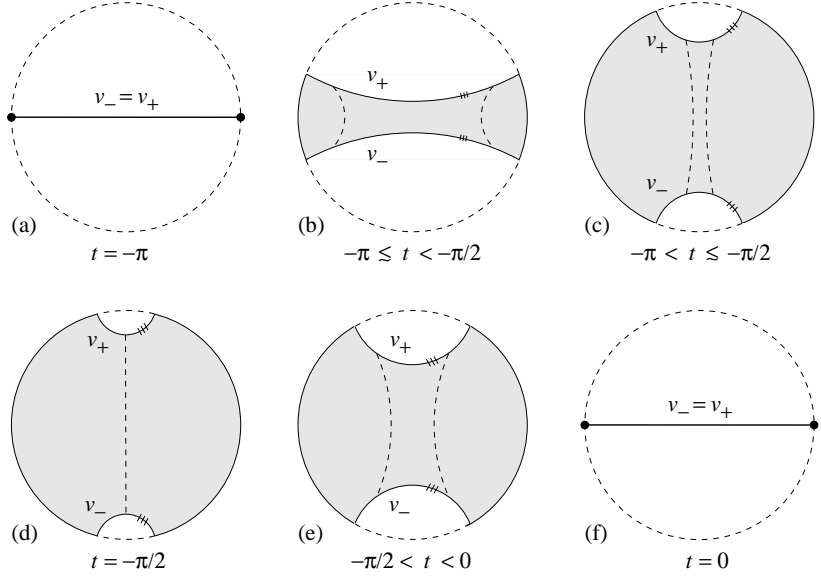}
\end{center}
\caption{The maximally extended BTZ black hole}
\label{bh0}
\end{figure}

Before considering the infalling particles, let us briefly describe the
maximally extended, matter free BTZ black hole in its rest frame
\cite{btz,adsbh}. It is a solution to the vacuum Einstein equations with
a negative cosmological constant, and it can be obtained by cutting and
gluing, very similar to the spacetime containing a point particle in
\sref{s-prt}. The only difference is that the holonomy $\uu$ has to be
spacelike. In the previous section we saw that the holonomy is related
to the mass $\imass$ of the black hole by \eref{btz-shell}, which states
that $\uu$ has to lie in a special conjugacy class of $\grpSL(2)$. A
simple group element with this property is
\begin{equation}\label{btz-hol}
    \uu = - e^{-\imass\,\ga_1} = - \cosh \imass \, \one + \sinh \imass \, \ga_1.
\end{equation}
The first question that arises is, are there any curves $\vdg_\pm$
inside the discs of constant $t$, such that one of them is mapped onto
the other by the isometry $\xx \mapsto \uu^{-1}\xx\,\uu$?  Such curves
do in fact exist. To find them, we can use the same symmetry argument
once again. The holonomy $\uu$ is an exponential of $\ga_1$, which means
that it is invariant under vertical reflections $\ga_2\mapsto-\ga_2$,
which is the same as $\ph\mapsto-\ph$ in anti-de-Sitter space. Assuming
the same for the curves $\vdg_+$ and $\vdg_-$, we make the ansatz that
the point $(t,r,\ph)\in\vdg_+$ is the image of the point
$(t,r,-\ph)\in\vdg_-$. Using the matrix representations \eref{ads-SL-rh}
for these points,
\begin{equation}\label{vv}
  \vv_\pm = \frac{1+r^2}{1-r^2} \, \om(t) 
          + \frac{2\, r}{1-r^2} \, \ga(\pm\ph),
\end{equation}
and evaluating the equation $\uu\,\vv_+=\vv_-\uu$, we arrive at the
following coordinate relations defining $\vdg_+$ and $\vdg_-$,
\begin{equation}\label{btz-faces}
   \vdg_\pm: \quad \frac{2\, r}{1+r^2} \, \sin \ph 
           = \mp \sin t \, \tanh\imass.   
\end{equation}
For $-\pi\le t\le0$, these curves are shown in \fref{bh0}. They are
again geodesics on the Poincar\'e disc, but their behaviour is very
different from that of the faces $\wdg_\pm$ for the lightlike particles.
The curves $\vdg_+$ and $\vdg_-$ do not intersect, except for $t=z\pi$,
$z\in\ZZ$, where they both coincide with the horizontal diameter of the
disc. This is the location of the fixed points.

If we define the space manifold to be the region between the two curves,
and identify the boundaries according to the isometry, then we obtain a
solution to the vacuum Einstein equations with a negative cosmological
constant, at least for times $t$ which are not multiples of $\pi$. At
$t=z\pi$, space degenerates to a line. As already mentioned in the end
of the previous section, it turns out that spacetime can be extended
beyond this singularity, but only if closed timelike curves are allowed.
If we do not allow them, or consider them as singularities as well, then
there is no way to continue spacetime beyond these lines. Let us
therefore restrict to a single time interval between two singularities,
say, the one with $-\pi\le t \le0$. There is then a future singularity
at $t=0$, and a past singularity at $t=-\pi$, very similar to those in
the case of a black and a white hole created by massive particles.

The global structure of the singularities is however somewhat different.
They are not located on half lines, emerging from the point where the
particles collide, and extending from there to $\scri$. Instead, both
singularities lie on full spacelike geodesics with two end points on
$\scri$. As a consequence, there are two distinct \emph{last points} on
$\scri$, the left and the right dot in \fref{bh0}(f), and also two
\emph{first points} on $\scri$, the dots in \fref{bh0}(a). Moreover, if
we look at the boundary of the disc more closely, we find that $\scri$
itself splits into two disconnected parts, the left and the right one,
each having its own first and last point. This is because the
identification of the curves $\vdg_+$ and $\vdg_-$ is such that the
right end of $\vdg_+$ is glued to the right end of $\vdg_-$, and the
left end of $\vdg_+$ to the left end of $\vdg_-$.

Hence, $\scri$ consists of two parts, both forming a closed circle. We
can also say that there are two spatial infinities. This situation is
well known, for example, from the maximally extended Schwarzschild black
hole. Indeed, the spacetime shown in \fref{bh0} has the same causal
structure. To see this, we have to consider the light cones emerging
from the first and the last points on $\scri$. They coincide pairwise,
because the time distance between the first and the last point is $\pi$,
which is exactly the time that a light ray needs to travel from one side
of the disc to the other. The actual construction of the light cones is
the same as in the previous section. The backward light cone of the last
point on, say, the right part $\scri$ is, at a time $t$, a geodesic
centered at $\cen=0$, with radius $\rad=-t$. As we need this later on,
let us write down a coordinate equation for this light cone. It is
obtained from \eref{origin-geo}, with $\cen=0$ and $\rad=-t$. This gives
\begin{equation}\label{hor-eq} 
  \frac{2\, r}{1+r^2} \, \cos \ph = \cos t.
\end{equation}
The backward light cone of the last point on the left part of $\scri$ is
the reflection of this at the vertical axis, $\ph\mapsto\pi-\ph$. Both
horizons are shown as dashed lines in \fref{bh0}. All together,
spacetime splits into four regions, and has the same global causal
structure as the maximally extended Schwarzschild black hole. We have
two external regions, to the left and to the right of both horizons,
which are causally completely disconnected. The region between the two
horizons for $t<-\pi/2$ is the interior of a white hole, from where
signals can be sent to both parts of $\scri$, and the region between the
horizons for $t>-\pi/2$ is the interior of a black hole, from where no
signal can be sent to either part of $\scri$.

The length of the horizon can be computed, most conveniently, in the
moment when the two horizons coincide in the middle of the disc, which
is the situation shown in \fref{bh0}(d). The radial coordinate $\rho$ of
the points where the horizon intersects with the curves $\vdg_\pm$ is
$\rho = \tanh(\imass/2)$, which follows from \eref{btz-faces} with
$\ph=\pm\pi/2$ and $t=-\pi/2$. The physical distance between these
points is
\begin{equation}
     \int\limits_{-\rho}^{\rho} \frac{2}{1-r^2} \, \dd r
      = 4 \, \arctanh \, \rho
      = 2 \imass.
\end{equation}
So, we recover the relation between the horizon length and the mass
$\imass$. A very different way to calculate the length of the horizon is
the following. Consider any two points in anti-de-Sitter space, and the
corresponding matrix representations $\xx,\yy\in\grpSL(2)$. If there is
a geodesic joining the two points, and only then the distance is
defined, then there is also a geodesic on the group manifold. Assume
that this geodesic is spacelike. There is then a unit spacelike vector
$\nn\in\algsl(2)$ such that $\yy=\xx\,e^{s\,\nn}$ for some positive
$s\in\RR$. This $s$ is the proper length of the geodesic joining $\xx$
and $\yy$. It is related to the matrices $\xx$ and $\yy$ by
\begin{equation}
   \ft12\Trr{\xx^{-1} \yy} = \cosh s.
\end{equation}
For a timelike distance, we have the same formula with $\cosh$ replaced
by $\cos$. Now, consider two corresponding points $(t,r,\ph)\in\vdg_+$
and $(t,r,-\ph)\in\vdg_-$, and their matrix representations \eref{vv}.
Using the formulas \eref{ga-om}, we find that
\begin{equation}\label{hor-len}
  \ft12\Trr{\vv_-\!\!^{-1}\vv_+} = 1 + 
  2 \, {\sin^2}\ph \, \Big(\frac{2\,r}{1-r^2}\Big)^2.
\end{equation}
This formula can be used to compute the horizon length at any time $t$.
What we need to do is to insert the coordinates of the intersection of
the curves \eref{btz-faces} with the horizon \eref{hor-eq}. The angular
coordinate can be found by dividing the two equations and using some
trigonometric identities, which gives
\begin{equation}
 \tan\ph = \tan t \, \tanh\imass \quad\Rightarrow\quad
   2 \, {\sin^2}\ph = 
   \frac{(1-\cos(2t))\,(\cosh(2\imass)-1)}{\cosh(2\imass)+\cos(2t)}.
\end{equation}
For the radial coordinate, we take the sum of the squares of the two
equations, and what we get is
\begin{equation}
  \Big(\frac{2\,r}{1+r^2}\Big)^2 =  
                   {\sin^2}t \,\, {\tanh^2}\imass + {\cos^2}t
  \quad\Rightarrow\quad
  \Big(\frac{2\,r}{1-r^2}\Big)^2 =  
        \frac{\cosh(2\imass)+\cos(2t)}{1-\cos(2t)}.
\end{equation}
If we insert this into \eref{hor-len}, the time dependence drops out,
and what remains is $\cosh(2\imass)$. This implies that the length of
the horizon is indeed independent of time, and equal to $2\imass$.

\subsection*{Infalling particles}
Let us now put in the particles. We can do this once again by cutting
and gluing, but this time we start from the BTZ black hole instead of
empty anti-de-Sitter space. We should expect that the particles somehow
cut away the second exterior region and also the white hole, because
these regions of spacetime did not occur in the previous section. If the
particles are falling in from, say, the right exterior region, then it
must be this region that remains. As an ansatz, let us assume, again
motivated by the symmetry of the picture, that one of the particles,
say, number two, is moving on the horizontal axis from the right to the
left, entering from $\scri$ at $t=-\pi/2$. Its holonomy is then given by
\begin{equation}\label{in-hol-2}
  \uu_2 = \one + \tan\eh_2 \, ( \ga_0 - \ga_1 ).
\end{equation}
Note that this is the same as the holonomy of the second particle in
\eref{two-hol}, so we can later apply some of the results from
\sref{s-col}. The parameter $\eh_2$ is determined by the following
condition. The holonomy $\uu_1$ of the other particle must be lightlike
as well, and the product of the two holonomies must be $\uu$. This gives
a relation between $\eh_2$ and $\imass$, namely
\begin{eqnarray}\label{eh-2-con}
   \ft12\Trr{\uu_1} = \ft12\Trr{\uu\,\uu_2\!^{-1}} = 
     \sinh\imass \, \tan\eh_2 + \cosh\imass = 1 
     \quad\Rightarrow\quad
     \tan\eh_2 = \coth(\imass/2).\qquad
\end{eqnarray}
There are then two possible choices for $\uu_1$, depending again on the
order in which the product is taken. If we make the ansatz
\begin{equation}\label{in-hol-1}
  \uu_{1\pm} = \one + \tan\eh_1 \, ( \ga_0 - \ga(\pm\dir) ) 
\end{equation}
and then solve the equations
\begin{equation}\label{hol-sec}
  \uu_{1-} \, \uu_2 = \uu , \qquad
  \uu_2 \, \uu_{1+} = \uu, 
\end{equation}
we find that the parameters $\dir$ and $\eh_1$, specifying the direction
and the energy of particle one, have to be chosen such that
\begin{equation}\label{in-par-def}
  \sin\dir = \tanh\imass, \qquad
  \tan\eh_1 = \cosh\imass \, \coth(\imass/2).
\end{equation}
But what does it mean that there are two possible choices for the
holonomy $\uu_1$? The answer is actually quite simple. Consider the
corresponding world line in the BTZ black hole spacetime, for $-\pi/2\le
t \le0$. It is the light ray defined by $r=-\tan(t/2)$ and
$\ph=\pm\dir$, depending on which sign we choose in \eref{in-hol-1}. As
can be easily checked, these world lines lie entirely inside the
surfaces $\vdg_\pm$, as defined in \eref{btz-faces}. They represent the
same world line, because the two surfaces are identified. 

\begin{figure}[t]
\begin{center}
  \epsfbox{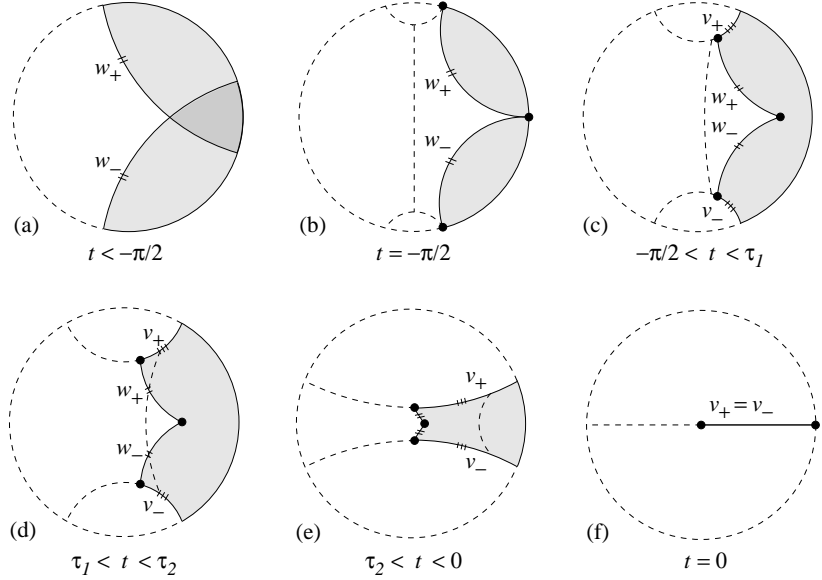}
\end{center}
\caption{The BTZ black hole with infalling particles}
\label{bh2}
\end{figure}

The resulting positions of the two particles as a function of time are
shown in \fref{bh2}(b--e). Particle one is always sitting at the
intersection of the curves $\vdg_\pm$ with the diameter of the disc
pointing into the angular direction $\pm\dir$.  Particle two is moving
on the horizontal diameter.  Both are at the same distance
$r=-\tan(t/2)$ from the center. They both enter from $\scri$ at
$t=-\pi/2$, and they collide in the center of the disc at $t=0$. To find
the wedges that the particles are cutting out from space, let us first
consider particle two, which is the same as the one already considered
in \sref{s-col}. There, we found that the wedge is bounded by the curves
\eref{wedge-eq-2},
\begin{equation}\label{wedge-eq-3}
   \wdg_\pm:\quad \frac{2\, r}{1+r^2} \, \sin(\eh_2\pm\ph) 
             =  - \sin t \, \sin \eh_2 . 
\end{equation}
Now, everything fits together very nicely. If we insert $r=-\tan(t/2)$
and $\ph=\pm\dir$ into these equation, we find that the world line of
particle one also lies inside the surfaces $\wdg_\pm$. Hence, particle
one always sits exactly at the intersection of the curves $\wdg_\pm$ and
$\vdg_\pm$, and $\wdg_\pm$ is the geodesic that joins the two particles.
If we cut out the wedge which is bounded by $\wdg_+$ and $\wdg_-$, the
space manifold becomes the shaded region shown in \fref{bh2}. Note that
it is now the wedge in front of the particle that is cut out, and not
the one behind. Furthermore, it is not necessary to cut out a second
wedge emerging from the other particle. This is already done by cutting
and gluing along the curves $\vdg_\pm$.

As all the identifications are provided by isometries of anti-de-Sitter
space, namely by the one represented by $\uu_2$ along $\wdg_\pm$, and by
that represented by $\uu$ along $\vdg_\pm$, it follows that the
resulting spacetime has a constant negative curvature everywhere, expect
on the world lines. One can also check that the holonomies of the
particles are the correct ones. For particle two this is obvious,
because in its neighbourhood spacetime looks exactly like the one
considered previously in \fref{pt1} or \ref{pt2}. To transport a
particle around particle one, we have to pass it once from $\vdg_-$ to
$\vdg_+$, which gives a Lorentz transformation $\uu$, and then from
$\wdg_+$ to $\wdg_-$, which contributes with a factor $\uu_2\!^{-1}$ to
the holonomy. The order in which the two factors have to be multiplied
depends on whether we start in the neighbourhood of the upper or the
lower representation of particle one on \fref{bh2}. In any case, we find
one of the holonomies $\uu_{1+}$ or $\uu_{1-}$, obeying \eref{hol-sec}.

What is then still missing is the continuation of spacetime to the past,
for $t<-\pi/2$. At the moment when the particles enter, the space
manifold is the shaded region of \fref{bh2}(b). This has the same shape
as the one in \fref{pt1}(d), which remained after a single massless
particle has passed through. Here, we have the reverse situation. It is
the space manifold before the particles enters. The fact that there are
actually two particles entering at the same time does not make any
difference, at least not at this moment, because both particles are
still outside the disc. We can continue to the past in the same way as
we continued to the future in \fref{pt1}. We let the curves $\wdg_\pm$
evolve further, according to \eref{wedge-eq-3}, and define the space
manifold to consist of the two partly overlapping shaded regions of
\fref{bh2}(a), glued together along their boundaries. From \sref{s-prt}
we know that this provides a skew foliation of anti-de-Sitter space.

All together, we have the same scenario as in the previous section,
except that the coordinates are slightly different. For $t<-\pi/2$, we
have an empty anti-de-Sitter space. At $t=-\pi/2$, two particles enter
from $\scri$. Shortly after that, in \fref{bh2}(c), there is still no
horizon, because the BTZ horizon lies still outside the shaded region.
In \fref{bh2}(d), one particle has fallen behind the horizon, but the
other particle is still outside. The time $\tau_1$ when the horizon
emerges at the position of the first particle can be found by setting
$r=-\tan(t/2)$ and $\ph=\dir$ in \eref{hor-eq}. The result is that
$\tan\tau_1 = -\cosh\imass$. After that, the horizon has the shape of a
closed loop with a cusp, pointing towards the second particle. It is
growing until the second particle also falls behind it at
$\tau_2=-\pi/4$, and thereafter the exterior region shown in
\fref{bh2}(e) looks exactly like one of the exterior regions of the
matter free BTZ black hole in \fref{bh0}.

Finally, the particles collide at $t=0$, inside the black hole, creating
the future singularity, which is shown in \fref{bh2}(f). It extends from the
point of collision in the center of the disc to $\scri$. This is what
remains from the space manifold. In contrast to the singularity in
\fref{bh0}(f), it is a half line only. It is the same spacelike geodesic
as the one on which the joint particle moves in \fref{bh1}(d--f). The
only difference is that now, in the rest frame of the black hole, it
entirely lies inside the disc of constant time at $t=0$. Therefore, we
do not see the tachyonic particle moving. Its world line is just there
at one moment of time. Nevertheless, the end point of the line on
$\scri$ is still the \emph{last point} on $\scri$, and the horizon is
the backward light cone of this point.
 
\vfill

\section*{Acknowledgements}

\noindent
Thanks to Ingemar Bengtsson, S\"oren Holst and Jorma Louko for many
discussions, and the Fysikum in Stockholm for hospitality.


\newpage

\begin{thebibliography}{99}
  
\bibitem{btz} M. Banados, C. Teitelboim, and J. Zanelli: The black
  hole in three dimensional space, \emph{Phys.\ Rev.\ Lett.}
  \textbf{69} (1992) 1849.
  
\bibitem{bhtz} M. Banados, M. Henneaux, C. Teitelboim, and J. Zanelli:
  Geometry of the 2+1 black hole, \emph{Phys.\ Rev.} \textbf{D48}
  (1993) 1506.
  
\bibitem{mannross} R. Mann and S. Ross: Gravitationally collapsing
  dust in (2+1)-dimensions, \emph{Phys.\ Rev.} \textbf{D47} (1993)
  3319
  
\bibitem{pp3d} S. Deser, R. Jackiw, and G. 't Hooft: Three dimensional
  Einstein gravity: dynamics of flat space, \emph{Ann.\ Phys.}
  \textbf{152} (1984) 220
  
\bibitem{thooft} G. 't Hooft: Quantization of point particles in
  (2+1)-dimensional gravity, \emph{Class.\ Quantum Grav.} \textbf{13}
  (1996) 1023.
  
\bibitem{matwel} H.-J. Matschull and M. Welling: Quantum mechanics of
  a point particle in 2+1 dimensional gravity, \emph{Class. Quantum
    Grav.} \textbf{15} (1998) 2981.
  
\bibitem{review} H.-J. Matschull: Three dimensional canonical quantum
  gravity, \emph{Class.\ Quantum Grav.} \textbf{12} (1995) 2621.
  
\bibitem{gott} J.R. Gott: Closed timelike curves produced by pairs of
  moving cosmic strings, \emph{Phys.\ Rev.\ Lett.} \textbf{66} (1991)
  1126.
  
\bibitem{hgott} M.P. Headrick and J.R. Gott: (2+1)-dimensional
  spacetimes containing closed timelike curves, \emph{Phys.\ Rev.}
  \textbf{D50} (1994) 7244.
  
\bibitem{cylco} S. \AA minneborg, I. Bengtsson, and S. Holst, and P.
  Peldan: Making Anti-de-Sitter black holes, \emph{Class.\ Quantum
    Grav.} \textbf{13} (1996) 2707.
  
\bibitem{adsbh} S. \AA minneborg, I. Bengtsson, D. Brill, S. Holst,
  and P. Peldan: Black holes and wormholes in 2+1 dimensions,
  
\bibitem{rotbh} S. \AA minneborg, I. Bengtsson, and S. Holst: A
  spinning anti-de-Sitter wormhole, gr-qc/9805028.

\end{thebibliography}
\end{document}